\documentclass[a4paper]{article}
\usepackage{jheppub}

\allowdisplaybreaks[2]

\newcommand\refr[1]     {ref.\,\cite{#1}}
\newcommand\refrs[1]    {refs.\,\cite{#1}}

\newcommand\eqn[1]     {eq.\,(\ref{#1})}
\newcommand\eqns[2]    {eqs.\,(\ref{#1}) and~(\ref{#2})}
\newcommand\eqnss[2]   {eqs.\,(\ref{#1})--(\ref{#2})}

\newcommand\fig[1]     {figure~{\ref{#1}}}

\newcommand\sect[1]    {section~{\ref{#1}}}
\newcommand\sects[2]   {sections~\ref{#1} and~\ref{#2}}
\newcommand\sectss[2]   {sections~\ref{#1}--\ref{#2}}

\newcommand\appx[1]     {appendix~\ref{#1}}

\def\beq{\begin{equation}}
\def\eeq{\end{equation}}
\def\bsp#1\esp{\begin{split}#1\end{split}}
\def\bal#1\eal{\begin{align}#1\end{align}}
\newcommand\nt         {\notag}

\newcommand\tsig[1]    {\sigma^{\rm{#1}}}
\newcommand\dsig[1]    {\rd\sigma^{{\rm #1}}}
\newcommand\dsiga[2]   {\rd\sigma^{{\rm #1,A}_{\scriptscriptstyle #2}}}

\newcommand\tgam[1]    {\Gamma^{\rm{#1}}}

\newcommand\dgam[2]    {\rd\Gamma^{{\rm #1}}_{#2}}
\newcommand\dgama[3]   {\rd\Gamma^{{\rm #1,A}_{\scriptscriptstyle #2}}_{#3}}

\newcommand\la         {\langle}
\newcommand\ra         {\rangle}
\newcommand\bra[3]     {\la {\cal M}_{#1}^{#2}#3|}
\newcommand\ket[3]     {|{\cal M}_{#1}^{#2}#3\ra}
\newcommand\braket[4]     {\la {\cal M}_{#1}^{#2}#3|{\cal M}_{#1}^{#4}#3\ra}
\newcommand\SME[3]     {|{\cal M}_{#1}^{#2}{#3}|^2}

\newcommand{\rd}       {{\rm{d}}}
\newcommand{\PS}[1]    {\rd\phi_{#1}}

\newcommand{\Y}[2]     {Y_{#1#2,Q}}

\newcommand{\cC}[2]    {{\cal C}_{#1}^{#2}}
\newcommand{\cS}[2]    {{\cal S}_{#1}^{#2}}
\newcommand{\cCS}[3]   {{\cal C}_{#1}^{~}{\cal S}_{#2}^{#3}}
\newcommand{\cSCS}[2]  {{\cal C}\kern-2pt{\cal S}_{#1}^{#2}}

\newcommand{\rC}       {{\rm C}}
\newcommand{\rS}       {{\rm S}}
\newcommand{\rSCS}     {{\rC}\kern-2pt{\rS}}
\newcommand{\IcC}[2]   {{\rC}_{#1}^{#2}}
\newcommand{\IcS}[2]   {{\rS}_{#1}^{#2}}

\newcommand{\IcSCS}[2] {\rC\kern-2pt\rS_{#1}^{#2}}
\newcommand{\bI}       {\bom{I}}

\newcommand{\Li}[2]   {\rm{Li}_{#1}\left(#2\right)}
\newcommand{\Log}[1]   {\ln #1}
\newcommand{\ze}[1]   {\zeta_{#1}}

\newcommand{\CF}       {C_{\rm{F}}}
\newcommand{\CA}       {C_{\rm{A}}}
\newcommand{\TR}       {T_{\rm{R}}}
\newcommand{\Nc}       {N_{\rm{c}}}
\newcommand{\Nf}       {\ensuremath{n_{\rm{f}}}}

\newcommand{\bT}       {\bom{T}}
\newcommand{\bTsq}[1]  {\bom{T}^2_{#1}}
\newcommand\qb         {{\bar q}}
\newcommand\bb         {{\bar {\rm b}}}

\newcommand{\fl}[1]   {#1}
\newcommand{\fla}[1]   {#1}

\def\AP{Altarelli--Parisi } 
\newcommand\bom[1]     {{\mbox{\boldmath $#1$}}}

\newcommand{\ep}       {\epsilon}
\newcommand{\Sep}       {S_\epsilon}
\newcommand{\Fep}       {S_\epsilon^{\MSbar}}

\newcommand{\cA}       {{\cal A}}

\newcommand\Oe[1]      {\ensuremath{\rm O(\ep^{#1})}}
\newcommand\Oa[1]      {\ensuremath{\rm O(\as^{#1})}}

\newcommand\ldot       {\!\cdot\!}
\newcommand\bbar      {\ensuremath{{\rm{b}\bar{\rm b}}}}
\newcommand\as  	       {\ensuremath{\alpha_{\rm{s}}}}
\newcommand\asR  	       {\ensuremath{\alpha_{\rm{s}}(\mu)}}
\newcommand\yb  	       {\ensuremath{y_{\rm b}}}
\newcommand\ybR  	       {\ensuremath{y_{\rm b}(\mu)}}
\newcommand\yc  	       {\ensuremath{y_{\rm cut}}}
\newcommand\EulerGamma	{\ensuremath{\gamma_{\rm{E}}}}
\newcommand\MSbar	  {\ensuremath{\overline{\rm{MS}}}}

\title{Higgs boson decay into b-quarks at NNLO accuracy}

\author[a]{Vittorio Del Duca,}
\author[b,c,1]{Claude Duhr,\note{On leave from the ``Fonds National de la Recherche Scientifique'' (FNRS), Belgium.}}
\author[d]{G\'abor Somogyi,}
\author[e]{Francesco Tramontano}
\author[d]{and Zolt\'an Tr\'ocs\'anyi}

\affiliation[a]{ 
Istituto Nazionale di Fisica Nucleare, 
Laboratori Nazionali di Frascati,\\ 
Via E. Fermi 40, I-00044 Frascati, Italy 
} 
\affiliation[b]{
PH Department, TH Unit, CERN, CH-1211 Geneva 23, Switzerland
}
\affiliation[c]{
Center for Cosmology, Particle Physics and Phenomenology (CP3),
Universit\'{e} Catholique de Louvain,
Chemin du Cyclotron 2, B-1348 Louvain-La-Neuve, Belgium
}
\affiliation[d]{ 
University of Debrecen and MTA-DE Particle Physics Research Group 
\\ H-4010 Debrecen, PO Box 105, Hungary
} 
\affiliation[e]{ 
Dipartimento di Fisica, Universit\`a degli studi di Napoli and 
INFN, Sezione di Napoli,\\ 
80125 Napoli, Italy
}

\emailAdd{Vittorio.DelDuca@lnf.infn.it}
\emailAdd{Claude.Duhr@cern.ch}
\emailAdd{Gabor.Somogyi@cern.ch}
\emailAdd{Zoltan.Trocsanyi@cern.ch}
\emailAdd{Francesco.Tramontano@cern.ch}

\abstract{
We compute the fully differential decay rate of the Standard Model
Higgs boson into b-quarks at next-to-next-to-leading order (NNLO) 
accuracy in $\as$. We employ a general subtraction scheme developed for 
the calculation of higher order perturbative corrections to QCD jet cross 
sections, which is based on the universal infrared factorization properties 
of QCD squared matrix elements. 
We show that the subtractions render the various contributions to the 
NNLO correction finite. In particular, we demonstrate analytically that the 
sum of integrated subtraction terms correctly reproduces the infrared poles 
of the two-loop double virtual contribution to this process.
We present illustrative differential distributions obtained by implementing 
the method in a parton level Monte Carlo program.
The basic ingredients of our subtraction scheme, used here for the first 
time to compute a physical observable, are universal and can be employed 
for the computation of more involved processes.
}


\arxivnumber{arXiv:1501.07226}

\preprint{CERN-PH-TH-2015-005, CP3-14-82}

\begin{document}
\maketitle
\flushbottom


\section{Introduction}
\label{sec:intro}

In run I, the ATLAS and CMS collaborations of the Large Hadron Collider
(LHC) discovered a new particle \cite{Aad:2012tfa,Chatrchyan:2012ufa}
with quantum numbers corresponding to those of the Higgs boson in the
Standard Model (SM) within the experimental accuracy of the measurements
\cite{Aad:2013xqa,Chatrchyan:2013iaa,Aad:2013wqa,Chatrchyan:2013mxa}.
Thus by now it is widely accepted that the new particle is 
the Higgs boson of the SM. Nevertheless, further more precise
measurements are being prepared for the upcoming run II. In particular,
a lot of emphasis is put on the precise determination of the couplings
of the Higgs boson to the heavy fermions to check whether the fermion 
masses are consistent with fermion mass generation in the SM.

Since the b-quark is quite light (its mass is only about 2\,\% of the 
vacuum expectation value of the Higgs field), the rate of associated
production of a b-quark pair with a Higgs boson is rather low. This fact, 
together with the overwhelming number of background events coming from 
direct QCD b-quark pair production makes the determination of the b-quark 
Yukawa coupling through $H\bbar$ production impossible. 
A better option that gives direct access to the $H\bbar$ Yukawa coupling 
is to measure the $H\to \bbar$ decay in the associated production of 
a Higgs boson with a $W$ or a $Z$ boson in a boosted or semi-boosted 
regime~\cite{Butterworth:2008iy}. In this scenario it is possible to 
use the kinematic and topological properties of the final states 
to isolate the $H\to \bbar$ decay. In this respect, first measurements
have been performed by the CMS~\cite{Chatrchyan:2013zna} 
and ATLAS~\cite{Aad:2014xzb} collaborations. 

Such search strategies may 
be aided by accurate modeling of QCD radiation in the $H\to \bbar$ decay, 
which motivates the computation of the fully differential decay rate at 
next-to-next-to-leading order (NNLO) accuracy in QCD perturbation theory. 
Computing fully differential cross sections and decay rates at NNLO turns 
out to be rather involved, however the last decade has witnessed substantial 
development \cite{Binoth:2000ps,Binoth:2004jv,Anastasiou:2003gr,
Anastasiou:2010pw,Weinzierl:2003fx,Weinzierl:2003ra,Catani:2007vq,
Campbell:1997hg,GehrmannDeRidder:2005cm,Daleo:2006xa,Daleo:2009yj,
Glover:2010im,Abelof:2011jv,Gehrmann:2011wi,GehrmannDeRidder:2012ja,Abelof:2012he,
Currie:2013vh,Czakon:2010td,Czakon:2011ve,Czakon:2014oma,
Boughezal:2011jf,Somogyi:2005xz,Somogyi:2006cz,Somogyi:2006da,Somogyi:2006db,
Somogyi:2008fc,Aglietti:2008fe,Somogyi:2009ri,Bolzoni:2009ye,Bolzoni:2010bt,
DelDuca:2013kw,Somogyi:2013yk} leading to a number of differential results 
for specific processes~\cite{Anastasiou:2003yy,Anastasiou:2003ds,Anastasiou:2004qd,
Anastasiou:2004xq,Anastasiou:2005qj,Anastasiou:2011qx,Melnikov:2006kv,
Weinzierl:2008iv,Weinzierl:2009nz,Weinzierl:2009ms,Weinzierl:2009yz,
Weinzierl:2010cw,Grazzini:2008tf,Catani:2009sm,Ferrera:2011bk,Catani:2011qz,
Grazzini:2013bna,Ferrera:2014lca,Cascioli:2014yka,Gehrmann:2014fva,
GehrmannDeRidder:2004tv,GehrmannDeRidder:2007jk,GehrmannDeRidder:2007hr,
Ridder:2013mf,Currie:2013dwa,Abelof:2014fza,Chen:2014gva,Abelof:2014jna,
Czakon:2013goa,Czakon:2014xsa,Gao:2012ja,Brucherseifer:2013iv,
Brucherseifer:2013cu,Boughezal:2013uia,Brucherseifer:2014ama}.

The first computation of the fully differential decay rate of the SM
Higgs boson into b-quarks at NNLO accuracy was published in
\refr{Anastasiou:2011qx}. That computation was performed with the
method of sector decomposition based on non-linear mappings
\cite{Anastasiou:2010pw}. Here we offer a different approach based on
the numerical implementation of the general subtraction scheme developed
in a series of papers for the computation of QCD jet cross sections at
NNLO accuracy~\cite{Somogyi:2005xz,Somogyi:2006cz,Somogyi:2006da,Somogyi:2006db,
Somogyi:2008fc,Aglietti:2008fe,Somogyi:2009ri,Bolzoni:2009ye,
Bolzoni:2010bt,DelDuca:2013kw,Somogyi:2013yk}. This method, which is used 
for the first time in this paper to compute a physical observable at NNLO, 
employs the universal infrared factorization of QCD squared matrix elements 
to define local subtraction terms for regulating the singularities
emerging in unresolved real radiation.

Specifically, we can write the NNLO correction to the cross section 
of a generic $m$-jet process as a sum of three contributions, the tree 
level double real radiation, the one-loop plus a single radiation, and 
the two-loop double virtual terms of the basic process under consideration,
\beq 
\tsig{NNLO} = 
	\int_{m+2}\dsig{RR}_{m+2} J_{m+2} 
	+ \int_{m+1}\dsig{RV}_{m+1} J_{m+1} 
	+ \int_m\dsig{VV}_m J_m\,,
\label{eq:sigmaNNLO} 
\eeq 
and rearrange it as follows,
\beq 
\tsig{NNLO} = 
	\int_{m+2}\dsig{NNLO}_{m+2} 
	+ \int_{m+1}\dsig{NNLO}_{m+1} 
	+ \int_m\dsig{NNLO}_m\,,
\label{eq:sigmaNNLOfin} 
\eeq 
where,
\bal
\dsig{NNLO}_{m+2} &= 
	\Big\{\dsig{RR}_{m+2} J_{m+2} 
	- \dsiga{RR}{2}_{m+2} J_{m} 
	-\Big[\dsiga{RR}{1}_{m+2} J_{m+1} 
	- \dsiga{RR}{12}_{m+2} J_{m}\Big]\Big\}_{\ep=0}\,, 
\label{eq:sigmaNNLOm+2} 
\\ 
\dsig{NNLO}_{m+1} &= 
	\Big\{\Big[\dsig{RV}_{m+1} 
	+ \int_1\dsiga{RR}{1}_{m+2}\Big] J_{m+1}  
	-\Big[\dsiga{RV}{1}_{m+1} 
	+ \Big(\int_1\dsiga{RR}{1}_{m+2}\Big)\strut^{{\rm A}_{\scriptscriptstyle 1}} 
	\Big] J_{m} \Big\}_{\ep=0}\,,
\label{eq:sigmaNNLOm+1} 
\\
\dsig{NNLO}_{m} &= 
	\Big\{\dsig{VV}_m 
	+ \int_2\Big[\dsiga{RR}{2}_{m+2} 
	- \dsiga{RR}{12}_{m+2}\Big] 
	+\int_1\Big[\dsiga{RV}{1}_{m+1} 
	+ \Big(\int_1\dsiga{RR}{1}_{m+2}\Big) \strut^{{\rm A}_{\scriptscriptstyle 1}} 
	\Big]\Big\}_{\ep=0} J_{m}\,.
\label{eq:sigmaNNLOm}
\eal
The subscripts on the integral signs are simply reminders that the
integration is over the phase space of $n = m$, $m+1$ or $m+2$ final
state particles. Above $J_n$ denotes the value of some infrared-safe 
observable $J$ evaluated on an $n$ parton final state.

The right-hand sides of \eqns{eq:sigmaNNLOm+2}{eq:sigmaNNLOm+1}
are integrable in four dimensions by construction
\cite{Somogyi:2005xz,Somogyi:2006cz,Somogyi:2006da,Somogyi:2006db},
while the integrability of \eqn{eq:sigmaNNLOm} in four dimensions is
ensured by the Kinoshita--Lee--Nauenberg (KLN) theorem on infrared-safe
quantities, provided that our subtraction scheme is well defined.

The counterterms which contribute to $\dsig{NNLO}_{m+2}$ and to
$\dsig{NNLO}_{m+1}$ were introduced in \refrs{Somogyi:2006da} and
\cite{Somogyi:2006db}. The integration of the real--virtual
counterterms (the last two terms of \eqn{eq:sigmaNNLOm}) was performed
in \refrs{Somogyi:2008fc,Aglietti:2008fe,Bolzoni:2009ye}. The
integral of the iterated single unresolved counterterm (the third term
of \eqn{eq:sigmaNNLOm}) was computed in \refr{Bolzoni:2010bt}. The
integration of the collinear-type contributions to the double unresolved
counterterm (the second term of \eqn{eq:sigmaNNLOm}) was performed in
\refr{DelDuca:2013kw}.  The soft-type contributions to the same
counterterm were presented in \refr{Somogyi:2013yk}. Most of these 
results were given as expansions in $\ep$ whose coefficients 
were computed numerically. Here we present the relevant integrals with
pole coefficients evaluated analytically, while the finite parts are
given numerically.  The final test on the consistency of our subtraction 
scheme is then to verify that \eqn{eq:sigmaNNLOm} is free of
singularities, as prescribed by the KLN theorem.  In this paper, we
perform that check analytically for the first time by computing the 
fully differential decay rate\footnote{In \eqnss{eq:sigmaNNLO}{eq:sigmaNNLOm} 
we presented the basic structure of our subtraction scheme for computing a 
generic cross section, however our method applies equally to decay rates, 
as spelled out in detail in \sectss{sec:LO}{sec:NNLO}.}
of the Higgs boson into b-quarks at NNLO.

The present work is the first physical application of this method,
therefore in order to facilitate reading we present the full
computation as implemented in a parton level Monte Carlo program
in detail. As usual in such codes, the jet function $J$ is computed
from generated momenta in $d=4$ dimensions, therefore, the
implementation of any infrared-safe physical quantity is
straightforward as demonstrated here.

The paper is organized as follows: in \sect{sec:notation}, the notation 
and conventions we use are introduced; in \sects{sec:LO}{sec:NLO}, we 
show the decay width at leading order and next-to-leading order (NLO) 
accuracy in $\as$; in \sect{sec:NNLO}, we display the counterterms and 
the insertion operators which are necessary to define the double real 
(\ref{eq:sigmaNNLOm+2}) and the real-virtual (\ref{eq:sigmaNNLOm+1}) 
contributions to the decay width, and we show that the double virtual 
contribution (\ref{eq:sigmaNNLOm}) is free of singularities; 
in \sect{sec:numres}, we show a selection of illustrative results;
we draw our conclusions in \sect{sec:concl}. The two appendices provide 
details on the matrix elements we use, as well as on the insertion operator 
used in the NLO computation.


\section{Notation}
\label{sec:notation}

We consider the partial decay width $\tgam{}_{H\to \bbar}[J]$ of the 
Higgs boson into a b-quark pair, for any infrared-safe observable $J$. 
Through NNLO in QCD, this decay width receives contributions from the 
following partonic subprocesses:
\\
\begin{center}\begin{tabular}{lll}
LO & $H(p_H) \to {\rm b}(p_1) + \bb(p_2)$ & tree level 
\\[0.5em]
NLO & $H(p_H) \to {\rm b}(p_1) + \bb(p_2) + g(p_3)$ & tree level
\\
	& $H(p_H) \to {\rm b}(p_1) + \bb(p_2)$ & one-loop
\\[0.5em]
NNLO & $H(p_H) \to {\rm b}(p_1) + \bb(p_2) + g(p_3) + g(p_4)$ & tree level
\\
	& $H(p_H) \to {\rm b}(p_1) + \bb(p_2) + q(p_3) + \qb(p_4)$ & tree level
\\
	& $H(p_H) \to {\rm b}(p_1) + \bb(p_2) + {\rm b}(p_3) + \bb(p_4)$ & tree level
\\
	& $H(p_H) \to {\rm b}(p_1) + \bb(p_2) + g(p_3)$ & one-loop
\\
	& $H(p_H) \to {\rm b}(p_1) + \bb(p_2)$ & two-loop
\end{tabular}\end{center}
~\\
where we show also the four-momenta of the particles in parentheses. 
We report the matrix elements corresponding to all subprocesses up to 
the required loop level in \appx{appx:MEs}.

We use the colour and spin space notation of ref.~\cite{Catani:1996vz} where 
the matrix element for a given subprocess, $\ket{n}{}{}$, is a vector in 
color and spin space, normalized such that the squared matrix element summed 
over colours and spins is given by
\beq
\SME{n}{}{} = \braket{n}{}{}{}\,,
\eeq
where $n$ is the number of particles in the final state.
The matrix element has the following formal loop expansion
\[
\ket{n}{}{} = \ket{n}{(0)}{} + \ket{n}{(1)}{} + \ket{n}{(2)}{} + \ldots\,,
\]
with the dots denoting higher-loop contributions. We will always
consider matrix elements computed in conventional dimensional
regularization (CDR) with \MSbar\ subtraction.  We will also use the
following $\otimes$ product notation to indicate the insertion of
colour charge operators between $\bra{}{(\ell_1)}{}$ and $\ket{}{(\ell_2)}{}$:
\beq
\bsp
\braket{}{(\ell_1)}{}{(\ell_2)} \otimes \bT_i \ldot\bT_k &\equiv
     \bra{}{(\ell_1)}{} \,\bT_i \ldot \bT_k \, \ket{}{(\ell_2)}{}\,,
\\[2mm]
\braket{}{(\ell_1)}{}{(\ell_2)}\otimes \{\bT_i \ldot \bT_k, \bT_j \ldot \bT_l\} &\equiv
     \bra{}{(\ell_1)}{} \{\bT_i \ldot \bT_k, \bT_j \ldot \bT_l\}
\ket{}{(\ell_2)}{}\,.
\esp
\label{eq:otimes-def}
\eeq
We use the customary
normalization of $\TR=1/2$ for the colour-charge operators, thus the
quadratic Casimirs are $\CA = 2\TR \Nc = \Nc$ in the adjoint and
$\CF = \TR(\Nc^2-1)/(\Nc) = (\Nc^2-1)/(2\Nc)$ in the fundamental
representation, where $\Nc=3$ is the number of colours.

The b-quark mass is much smaller than the scale of the problem that 
is the Higgs boson mass, therefore, we treat the b-quarks as massless, 
both in the matrix elements and phase space integrals, retaining the 
b-quark mass only in the Yukawa coupling. We neglect the t-quark 
throughout and consider $\Nf=5$ light quark flavours.

In QCD the renormalized amplitudes are obtained from the unrenormalized
ones by replacing the bare couplings $\yb^B$ and $\as^B$ with their
renormalized counterparts evaluated at the renormalization scale $\mu$
\bal
\yb^B\,\mu_0^\ep &= \yb\,\mu^\ep \bigg\{
	1
	-\frac{\as}{4\pi} \frac{3\CF}{\ep}
	+\bigg(\frac{\as}{4\pi}\bigg)^2
		\bigg[
			\bigg(\frac{11\CA}{2} + \frac{9\CF}{2} - 2\Nf \TR\bigg)\frac{1}{\ep^2}
\nt \\&\qquad\qquad\qquad\qquad\qquad\qquad\qquad
			-\bigg(\frac{97\CA}{12} + \frac{3\CF}{4} - \frac{5\Nf \TR}{3}\bigg)
			\frac{1}{\ep}\bigg] + \Oa{3}\bigg\}\,,
\label{eq:ybRenorm}
\\
\as^B \mu_0^{2\ep} &= \frac{\as}{\Fep} \mu^{2\ep}
	\bigg[1 - \frac{\as}{4\pi}\frac{\beta_0}{\ep} + \Oa{2}\bigg]\,,
\label{eq:asRenorm}
\eal
where
\beq
\beta_0 = \frac{11\CA}{3} - \frac{4\Nf \TR}{3}\,,
\eeq
and $\Fep = (4\pi)^\ep \exp(-\ep \EulerGamma)$ corresponds to \MSbar\
subtraction. Although the factor $(4\pi)^\ep \exp(-\ep \EulerGamma)$
is often abbreviated as $\Sep$ in the literature, we reserve the latter
to denote 
\beq
\Sep = \frac{(4\pi)^\ep}{\Gamma(1-\ep)}\,.
\eeq
On the right-hand side, $\yb\equiv\ybR$ and $\as\equiv\asR$ are the 
dimensionless renormalized couplings in the \MSbar\ scheme 
evaluated at the renormalization scale $\mu$. 

The $n$ particle massless phase space measure reads
\beq
\PS{n}(Q^2) \equiv \PS{n}(p_1,\ldots,p_n;Q) = 
	\Bigg[\prod_{i=1}^{n}\frac{\rd^d p_i}{(2\pi)^{d-1}}
		\delta_{+}(p_i^2)\Bigg]
	(2\pi)^d \delta^{(d)}(p_1 + \ldots + p_n - Q)\,.
\eeq
Throughout the paper, we will use $y_{ik}$ to denote twice the
dot-product of two momenta, scaled by the total momentum squared $Q^2$.
For example,
\beq
y_{ik} = \frac{2p_i\cdot p_k}{Q^2}\qquad\mbox{and}\qquad
y_{iQ} = \frac{2p_i\cdot Q}{Q^2}\,.
\label{eq:yik-def}
\eeq
We also introduce the combination
\beq
Y_{ik,Q} = \frac{y_{ik}}{y_{iQ} y_{kQ}}
\label{eq:YikQ-def}
\eeq
for later convenience.


\section{Leading order}
\label{sec:LO}

Let us denote the Born differential decay rate by,
\beq
\dgam{B}{2} = \frac{1}{2m_H} \PS{2}(m_H^2)\, \SME{\bbar}{(0)}{}\,.
\label{eq:dgamLO_2}
\eeq
Then the leading order decay width is,
\beq
\tgam{B}[J] = 
	\int_2 \dgam{B}{2} J_2 = 
	\frac{1}{2m_H} \int \PS{2}(m_H^2)\, \SME{\bbar}{(0)}{} J_2\,.
\eeq
Here $J$ is an infrared-safe observable whose value evaluated on a kinematic 
configuration with two partons is $J_2$.
For the inclusive decay width ($J\equiv 1$) at leading order we 
have
\beq
\tgam{LO} = \tgam{B}[J=1] = \frac{\yb^2 m_H \Nc}{8\pi}\,,
\label{eq:GB}
\eeq
where the expression on the right-hand side is the four-dimensional result. 


\section{Next-to-leading order}
\label{sec:NLO}

%
%

\subsection{Real emission contribution}
\label{sec:NLO-R}

The real emission contribution to the differential decay width reads
\beq
\dgam{R}{3} = 
	\frac{1}{2m_H} \PS{3}(m_H^2)\, \SME{\bbar g}{(0)}{}\,.
\label{eq:GRdiff}
\eeq
$\dgam{R}{3}$ is divergent when the radiated gluon becomes unresolved
(soft, or collinear with one of the b-quarks). In order to regularize
it, we subtract an approximate decay rate,
\beq
\dgama{R}{1}{3} =
	\frac{1}{2m_H} \PS{3}(m_H^2)\, \cA_1 \SME{\bbar g}{(0)}{}\,,
\label{eq:GRdiffA1}
\eeq
where the counterterm for processes with $m+1$ partons in the final state is 
given by \cite{Somogyi:2006cz,Somogyi:2006da},
\beq
\cA_1 \SME{m+1}{(0)}{} = 
	\sum_{r=1}^{m+1} 
	\Bigg[ \sum_{\substack{i=1 \\ i\ne r}}^{m+1} \frac1{2} \cC{ir}{(0,0)}
	- \Bigg( \cS{r}{(0,0)} 
		- \sum_{\substack{i=1 \\ i\ne r}}^{m+1} \cC{ir}{}\cS{r}{(0,0)} \Bigg) 
\Bigg] \,.
\label{eq:A1def}
\eeq
In \eqn{eq:A1def} the functions $\cC{ir}{(0,0)}$ and $\cS{r}{(0,0)}$
appearing in the right-hand side correspond to counterterms which
regularize the $p_i||p_r$ collinear limit and the $p_r\to 0$ soft
limit. In order to avoid double counting in the overlapping
soft-collinear region, we must add back a soft-collinear counterterm,
$\cC{ir}{}\cS{r}{(0,0)}$. The precise definitions of these subtractions
are given in \refrs{Somogyi:2006cz,Somogyi:2006da}. In our convention
the indices of $\cC{ir}{(0,0)}$ are not ordered, $\cC{ir}{(0,0)} =
\cC{ri}{(0,0)}$. Since the sums over $i$ and $r$ in \eqn{eq:A1def} are
likewise not ordered, the factor of $\frac12$ assures that we count
each collinear limit precisely once. Finally, the superscript $(\ell_1,\ell_2)$
means that the corresponding counterterm involves the product (in
colour or spin space) of an $\ell_1$-loop unresolved kernel (an \AP
splitting function or a soft eikonal current) with an $\ell_2$-loop squared
matrix element. Thus, $(0,0)$ means that we consider a tree level
collinear or soft function acting on a tree level reduced matrix
element.  Such superscripts will appear also for other counterterms
throughout the paper.  For definitiveness, we spell out \eqn{eq:A1def}
explicitly for $H \to \bbar g$ ($m=2$) below,
\beq
\cA_1 \SME{\bbar g}{(0)}{} = 
	\cC{13}{(0,0)} + \cC{23}{(0,0)} + \cS{3}{(0,0)} 
	- \cC{13}{}\cS{3}{(0,0)} - \cC{23}{}\cS{3}{(0,0)}\,,
\eeq
where the b, $\bb$ and gluon carry the labels 1, 2 and 3.

With the counterterms given in \refrs{Somogyi:2006cz,Somogyi:2006da} it 
is straightforward to check that the difference 
\beq
\dgam{NLO}{3}\equiv \dgam{R}{3} J_3 - \dgama{R}{1}{3} J_2
\label{eq:dgamNLO_3}
\eeq
is integrable in all kinematic limits. 
Then, the regularized real contribution to the decay rate,
\beq
\tgam{NLO}_3[J] = \int_3 
\big[\dgam{NLO}{3}\big]_{\ep=0}
\eeq
is finite in four dimensions for any infrared-safe observable. An explicit 
calculation for the contribution to the total decay width from the real 
emission part plus subtractions yields
\beq
\tgam{NLO}_3[J=1] = 
	\tgam{LO} \frac{\as}{\pi} \CF \frac{1729}{450}\,.
\label{eq:GNLO3}
\eeq

%
%

\subsection{Virtual contribution}
\label{sec:NLO-V}

The virtual contribution to the differential decay width reads
\beq
\dgam{V}{2} = 
	\frac{1}{2m_H} \PS{2}(m_H^2)\, 
	2\Re \braket{\bbar}{(0)}{}{(1)}\,,
\label{eq:GVdiff}
\eeq
and is of course divergent in four dimensions. Its $\ep$-expansion reads
(see \eqn{eq:bb1loop})
\beq
\dgam{V}{2} = 
	\dgam{B}{}
	\frac{\as}{2\pi}\frac{\Sep}{\Fep}
	\bigg(\frac{\mu^2}{m_H^2}\bigg)^\ep \CF
	\left[
		- \frac{2}{\ep^2} 
		- \frac{3}{\ep}
		- 2 + \pi^2 + 3 L
		+ \Oe{} \right]\,,
\label{eq:GVdiff-exp}
\eeq
where we have introduced the abbreviation 
$L=\Log{\left(\frac{\mu^2}{m_H^2}\right)}$.
In \eqn{eq:GVdiff-exp}, $\dgam{B}{}$ denotes the $d$-dimensional Born
decay rate as given in \eqn{eq:dgamLO_2}.  

By the KLN theorem, the integral of the approximate decay rate
precisely cancels the divergences of the virtual piece, so adding back
what we have subtracted from the real correction, the virtual contribution 
becomes finite as well. We have performed the integration of
the  various subtraction terms analytically in \refr{Somogyi:2006cz}
and here we only quote the result, which can be written as,
\beq
\int_1\dgama{R}{1}{m+1} = 
	\dgam{B}{m} \otimes \bI_1^{(0)}(\{p\}_m;\ep)\,,
\label{eq:INTGRdiffA1}
\eeq
where the $\otimes$ product is defined in \eqn{eq:otimes-def} and the
insertion operator is in general given by \cite{Somogyi:2006cz}%
\footnote{The expansion parameter in \refr{Somogyi:2006cz} was chosen
$\as/\Fep$ implicitly, with the harmless factor $1/\Fep$ suppressed.
For the sake of clarity we reinstate the factor $1/\Fep$ here, as well
as in all other insertion operators in eqns.\ (\ref{eq:I2}),
(\ref{eq:I12}), (\ref{eq:I11-bare}) and (\ref{eq:I1100}) below.}
\beq
\bI_1^{(0)}(\{p\}_m;\ep) = 
	\frac{\as}{2\pi}\frac{\Sep}{\Fep}
	\bigg(\frac{\mu^2}{Q^2}\bigg)^\ep
	\sum_{i=1}^m\bigg[\IcC{1,i}{(0)}(y_{iQ};\ep) \bT_i^2 
		+ \sum_{\substack{k=1 \\ k\ne i}}^m 
		\IcS{1}{(0),(i,k)}(Y_{ik,Q};\ep) \bT_i \bT_k\bigg]\,.
\label{eq:I10}
\eeq
The variables $y_{iQ}$ and $Y_{ik,Q}$ were defined in \eqns{eq:yik-def}{eq:YikQ-def}
and $Q^\mu$ is the total incoming momentum.
The functions $\IcC{1,i}{(0)}(y_{iQ};\ep)$ and $\IcS{1}{(0),(i,k)}(Y_{ik,Q};\ep)$ 
have been computed as Laurent expansions in $\ep$ in \refr{Somogyi:2006cz} 
and are recalled here up to finite terms in \appx{appx:I103j}. 
We mention that there is no one-to-one correspondence between the unintegrated 
subtraction terms in \eqn{eq:A1def} and the kinematic functions that appear 
in \eqn{eq:I10}. The latter are obtained from the former after summing over all 
unobserved quantum numbers (colour and flavour) in addition to integrating over 
the unresolved momentum, and organizing the result in colour and flavour space. 
Loosely speaking, the integrated form of $\cC{ir}{(0)}$ enters $\IcC{1,i}{(0)}$
and that of $\cS{r}{(0)}$ enters $\IcS{1}{(0),(i,k)}$. However, we are free to 
assign the integrated form of $\cC{ir}{}\cS{r}{(0)}$ to either of the 
integrated counterterms and this final organization was performed differently 
in \refr{Somogyi:2006cz} and in this paper. In \refr{Somogyi:2006cz}, the 
integrated form of $\cC{ir}{}\cS{r}{(0)}$ was grouped into $\IcS{1}{(0),(i,k)}$, 
while here we find it more convenient to group it into 
$\IcC{1,i}{(0)}$.

For $H \to \bbar$, with only two partons in the final state the colour 
connections factorize completely,
\beq
\bT_1\bT_2 = - \CF\,.
\label{eq:connect2}
\eeq
Furthermore, momentum conservation implies that
\beq
y_{1Q} = y_{2Q} = Y_{12,Q} = y_{12} = 1\,.
\label{eq:Bornkin}
\eeq
Thus, the insertion operator $\bI_1^{(0)}$ becomes,
\beq
\bI_1^{(0)}(p_1,p_2;\ep) = 
	\frac{\as}{2\pi} \frac{\Sep}{\Fep}
	\left(\frac{\mu^2}{m_H^2}\right)^\ep
	2\CF\bigg[
	\IcC{1,q}{(0)}(1;\ep) 
	-\IcS{1}{(0),(1,2)}(1;\ep)\bigg]\,,
\label{eq:I10-2jet}
\eeq
where, as indicated, we must evaluate all functions with arguments equal to one. 
The Laurent expansion of \eqn{eq:I10-2jet} in $\ep$ is, 
\beq
\bsp
\bI_1^{(0)}(p_1,p_2;\ep) &= 
	\frac{\as}{2\pi} \frac{\Sep}{\Fep}
	\bigg(\frac{\mu^2}{m_H^2}\bigg)^\ep 
\\&\quad\times 
	\CF
	\left[
		\frac{2}{\ep^2} 
		+ \frac{3}{\ep}
		+ \frac{1267}{450} - \pi^2
		+ \left( \frac{137 \pi^2}{90} - \frac{707519}{13500} \right) \ep
		- 95.9144 \ep^2
		+ \Oe{3} \right]\,,
\label{eq:I10-2jetexp}
\esp
\eeq
where, for future reference, we have also provided the $\Oe{}$ part in
terms of rational numbers and known transcendental constants. 
The uncertainty of the $\Oe{2}$ numerical result, as well as those of all 
other numerical results we show affect the last quoted digit, unless 
specifically stated otherwise.

It is easy to check that the expression
\beq
\dgam{NLO}{2} \equiv 
	\bigg[\dgam{V}{2} + \int_1 \dgama{R}{1}{3}\bigg] J_2\,,
\label{eq:dgamNLO_2}
\eeq
is free of $\ep$-poles. Hence
\beq
\tgam{NLO}_2[J] = \int_2
\big[\dgam{NLO}{2}\big]_{\ep=0}
\eeq
is finite in four dimensions for any infrared-safe observable. For the 
contribution to the total width from the virtual part plus integrated 
subtractions we find
\beq
\tgam{NLO}_2[J=1] =
	\tgam{LO}\,\frac{\as}{\pi}
	\bigg(\frac{367}{900}\CF + \frac{3}{2}\CF L \bigg)\,.
\label{eq:GNLO2}
\eeq
Combining \eqns{eq:GNLO3}{eq:GNLO2}, we obtain the full NLO correction to the 
total decay rate, 
\beq
\tgam{NLO} = \tgam{NLO}_3[J=1] + \tgam{NLO}_2[J=1] =   
	\tgam{LO}\,\frac{\as}{\pi}
	\bigg(\frac{17}{4}\CF + \frac{3}{2}\CF\,L\bigg)\,.
\label{eq:tgamNLO}
\eeq
As $\CF=\frac{4}{3}$ in the conventions used, we recover the 
well-known NLO result~\cite{Gorishnii:1990zu,Gorishnii:1991zr,Baikov:2005rw}.


\section{Next-to-next-to-leading order}
\label{sec:NNLO}

%
%

\subsection{Double real emission contribution}
\label{ssec:NNLO-RR}

The double real emission contribution to the differential decay width is
\beq
\dgam{RR}{4} = 
	\frac{1}{2m_H} \PS{4}(m_H^2)\,
	 \Bigg( 
	 \frac{1}{2!} \SME{\bbar gg}{(0)}{} 
	 + \sum_{q\ne {\rm b}} \SME{\bbar q{\bar q}}{(0)}{} 
	 + \frac{1}{(2!)^2} \SME{\bbar \bbar}{(0)}{} \Bigg) \,,
\label{eq:GRRdiff}
\eeq
and its integral over the phase space is divergent in four dimensions
due to kinematic singularities emerging in unresolved regions. In order
to regularize the singularities of \eqn{eq:GRRdiff} due to two unresolved
partons, we subtract an approximate decay rate,
\beq
\dgama{RR}{2}{4} = 
	\frac{1}{2m_H} \PS{4}(m_H^2)\, 
		\Bigg( 
		\frac{1}{2!} \cA_2 \SME{\bbar gg}{(0)}{} 
		+ \sum_{q\ne {\rm b}} \cA_2 \SME{\bbar q{\bar q}}{(0)}{} 
		+ \frac{1}{(2!)^2} \cA_2 \SME{\bbar \bbar}{(0)}{} \Bigg) \,,
\label{eq:GRRA2diff}
\eeq
where the double unresolved counterterm for processes with $m+2$ partons in 
the final state is \cite{Somogyi:2006da}
\beq
\bsp
\cA_2 \SME{m+2}{(0)}{} &=
\sum_{r=1}^{m+2}\sum_{s=1}^{m+2}\Bigg\{
\sum_{\substack{i=1 \\ i\ne r,s}}^{m+2}\Bigg[\frac16\, \cC{irs}{(0,0)}
+ \sum_{\substack{j=1 \\ j\ne i,r,s}}^{m+2} \frac18\, \cC{ir;js}{(0,0)}
\\ 
&\quad +\, \frac12\,\Bigg( \cSCS{ir;s}{(0,0)}
- \cC{irs}{}\cSCS{ir;s}{(0,0)} 
- \sum_{\substack{j=1 \\ j\ne i,r,s}}^{m+2} \cC{ir;js}{} \cSCS{ir;s}{(0,0)} \Bigg)
\\ 
&\quad -\,\, \cSCS{ir;s}{}\cS{rs}{(0,0)}
- \frac12\, \cC{irs}{}\cS{rs}{(0,0)}
+ \cC{irs}{}\cSCS{ir;s}{}\cS{rs}{(0,0)}
\\ 
&\quad + \sum_{\substack{j=1 \\ j\ne i,r,s}}^{m+2} 
	\frac12\, \cC{ir;js}{}\cS{rs}{(0,0)}\Bigg] + \frac12\, \cS{rs}{(0,0)}
\Bigg\}
\,.
\label{eq:RR_A2}
\esp
\eeq
In \eqn{eq:RR_A2}, the functions $\cC{irs}{(0,0)}$, $\cC{ir;js}{(0,0)}$, 
$\cSCS{ir;s}{(0,0)}$ and $\cS{rs}{(0,0)}$ denote counterterms which regularize 
the $p_i||p_r||p_s$ triple collinear, the $p_i||p_r$, $p_j||p_s$ double collinear, 
the $p_i||p_r$, $p_s\to 0$ one collinear, one soft (collinear+soft) and the
$p_r\to 0$, $p_s\to 0$ double soft limits. The rest of the counterterms which appear in 
\eqn{eq:RR_A2} account for the double or triple overlap of limits, their role is 
to make sure that no multiple subtractions are performed in overlapping double 
unresolved regions. Thus, for instance, $\cC{irs}{}\cSCS{ir;s}{(0,0)}$ accounts for the 
triple collinear limit of the collinear+soft counterterm, and the rest of the 
counterterms have a similar interpretation as suggested by the notation. The 
precise definitions of all functions appearing in \eqn{eq:RR_A2} were given in 
\refr{Somogyi:2006da}. As in our convention the collinear indices of 
counterterms and the sums over them in \eqn{eq:RR_A2} are not ordered, the 
factors of $\frac16$, $\frac18$, etc., are needed so that each limit is 
counted precisely once.

After subtracting the double unresolved approximate cross section, the 
difference
\beq
\dgam{RR}{4} - \dgama{RR}{2}{4}
\eeq
is however still singular in the single unresolved regions of phase space. 
To regularize it, we also subtract
\beq
\dgama{RR}{1}{4} = 
	\frac{1}{2m_H} \PS{4}(m_H^2)\, 
		\Bigg( 
		\frac{1}{2!} \cA_1 \SME{\bbar gg}{(0)}{} 
		+ \sum_{q\ne b} \cA_1 \SME{\bbar q{\bar q}}{(0)}{} 
		+ \frac{1}{(2!)^2} \cA_1 \SME{\bbar \bbar}{(0)}{} \Bigg)\,,
\label{eq:GRRA1diff}
\eeq
where $\cA_1$ has been defined in \eqn{eq:A1def}.  To avoid double
subtraction in overlapping single and double unresolved regions of
phase space, we must also consider
\beq
\dgama{RR}{12}{4} = 
	\frac{1}{2m_H} \PS{4}(m_H^2)\,
	\Bigg( 
		\frac{1}{2!} \cA_{12} \SME{\bbar gg}{(0)}{} 
		+ \sum_{q\ne b} \cA_{12} \SME{\bbar q{\bar q}}{(0)}{} 
		+ \frac{1}{(2!)^2} \cA_{12} \SME{\bbar \bbar}{(0)}{} \Bigg) \,.
\label{eq:GRRA12diff}
\eeq
The general formula for the iterated single unresolved counterterm is
\beq
\cA_{12} \SME{m+2}{(0)}{} =
\sum_{t=1}^{m+2}\Bigg[
\sum_{\substack{k=1 \\ k\ne t}}^{m+2} \frac12 \,\cC{kt}{} \cA_2 \SME{m+2}{(0)}{}
+\Bigg(\cS{t}{} \cA_2 \SME{m+2}{(0)}{} -
\sum_{\substack{k=1 \\ k\ne t}}^{m+2} \cC{kt}{}\cS{t}{} \cA_2 \SME{m+2}{(0)}{} \Bigg) 
\Bigg]\,,
\label{eq:RR_A12}
\eeq
where the three terms above are given by~\cite{Somogyi:2006da},
\bal
\cC{kt}{} \cA_2 & = 
\sum_{\substack{r=1 \\ r\ne k,t}}^{m+2} 
\Bigg[\cC{kt}{}\cC{ktr}{(0,0)} +\cC{kt}{}\cSCS{kt;r}{(0,0)} 
- \cC{kt}{}\cC{ktr}{}\cSCS{kt;r}{(0,0)}
- \cC{kt}{}\cC{rkt}{}\cS{kt}{(0,0)}
\nt\\ &\qquad\qquad
+ \sum_{\substack{i = 1 \\ i\ne r,k,t}}^{m+2}\Bigg(
\frac12\,\cC{kt}{}\cC{ir;kt}{(0,0)} 
- \cC{kt}{}\cC{ir;kt}{}\cSCS{kt;r}{(0,0)}\Bigg)\Bigg]
+ \cC{kt}{}\cS{kt}{(0,0)} \,,
\label{eq:CktA2}
\\
\cS{t}{} \cA_2 & =
\sum_{\substack{r=1 \\ r\ne t}}^{m+2}\Bigg\{\sum_{\substack{i = 1 \\ i\ne r,t}}^{m+2} 
\Bigg[\frac12\Bigg(\cS{t}{}\cC{irt}{(0,0)} +\cS{t}{}\cSCS{ir;t}{(0,0)} 
- \cS{t}{}\cC{irt}{}\cSCS{ir;t}{(0,0)}\Bigg)
\nt\\ &\qquad\qquad\qquad
- \cS{t}{}\cC{irt}{}\cS{rt}{(0,0)} - \cS{t}{}\cSCS{ir;t}{}\cS{rt}{(0,0)} 
+ \cS{t}{}\cC{irt}{}\cSCS{ir;t}{}\cS{rt}{(0,0)}\Bigg] 
+ \cS{t}{}\cS{rt}{(0,0)}\Bigg\} \,,
\label{eq:StA2}
\\
\cC{kt}{}\cS{t}{} \cA_2 & =
  \sum_{\substack{r=1 \\ r\ne k,t}}^{m+2} \Bigg[\cC{kt}{}\cS{t}{}\cC{krt}{(0,0)} 
+ \sum_{\substack{i = 1 \\ i\ne r,k,t}}^{m+2}\Bigg(
\frac12\cC{kt}{}\cS{t}{}\cSCS{ir;t}{(0,0)} 
- \cC{kt}{}\cS{t}{}\cSCS{ir;t}{}\cS{rt}{(0,0)}\Bigg)
\nt\\ &\qquad\qquad
- \cC{kt}{}\cS{t}{}\cC{krt}{}\cS{rt}{(0,0)}
- \cC{kt}{}\cS{t}{}\cC{rkt}{}\cS{kt}{(0,0)}
+ \cC{kt}{}\cS{t}{}\cS{rt}{(0,0)}\Bigg]
+ \cC{kt}{}\cS{t}{}\cS{kt}{(0,0)}\,.
\label{eq:CktStA2}
\eal
The interpretation of the various terms in \eqnss{eq:CktA2}{eq:CktStA2} are 
suggested by the notation: for instance, $\cC{kt}{}\cC{ktr}{(0,0)}$ in \eqn{eq:CktA2} 
accounts for the $p_k||p_t$ single collinear limit of the $\cC{ktr}{(0,0)}$ 
triple collinear counterterm, while, for example, $\cS{t}{}\cC{irt}{(0,0)}$ in 
\eqn{eq:StA2} represents the counterterm appropriate to the $p_t\to 0$ soft 
limit of $\cC{irt}{(0,0)}$. 
Thus, $\cA_{12} \SME{m+2}{(0)}{}$ cancels the single unresolved singularities 
of the double unresolved subtraction term $\cA_{2} \SME{m+2}{(0)}{}$. However, 
very importantly, it can also be shown \cite{Somogyi:2006da} that 
$\cA_{12} \SME{m+2}{(0)}{}$ simultaneously cancels the double unresolved 
singularities of the single unresolved subtraction term $\cA_{1} \SME{m+2}{(0)}{}$ 
and so properly accounts for the overlap of single and double unresolved subtractions.
All of the counterterms appearing in \eqnss{eq:CktA2}{eq:CktStA2} were precisely 
defined in \refr{Somogyi:2006da}. As before, the collinear indices and sums over 
them in \eqnss{eq:RR_A12}{eq:CktStA2} are not ordered, hence the appearance of the 
factors of $\frac12$ at various instances.

With these definitions, the difference
\beq
\dgam{NNLO}{4} \equiv
	\dgam{RR}{4} J_4 - \dgama{RR}{2}{4} J_2 
	- \dgama{RR}{1}{4} J_3 + \dgama{RR}{12}{4} J_2
\label{eq:dgamNNLO_4}
\eeq
can be shown to be integrable in all kinematic limits~\cite{Somogyi:2006da}. 
Thus, the regularized double real contribution to the decay rate
\beq
\tgam{NNLO}_4[J] = \int_4 
\big[\dgam{NNLO}{4}\big]_{\ep=0}
\label{eq:tgammaNNLO_4}
\eeq
is finite in four dimensions for any infrared-safe observable and can
be computed with standard numerical techniques.
For the total cross section ($J=1$) at $\mu=m_H$ ($L=0$) we find,
\beq
\tgam{NNLO}_4[J=1] =
     \tgam{LO}\,\left(\frac{\as}{\pi}\right)^2
     1.05(1)\,.
\label{eq:tgamNNLO_4}
\eeq
This numerical value has been obtained by implementing \eqn{eq:tgammaNNLO_4} 
in a fully differential parton level Monte Carlo program using four 
dimensional double real emission matrix elements and phase space. 
However, we have also reproduced the result by integrating the matrix 
elements and subtraction terms directly in $d$ dimensions and then summing 
the separate contributions. We stress that this is a highly non-trivial 
cross check, as both calculations are very different conceptually and technically.

%
%

\subsection{Real--virtual contribution}
\label{ssec:NNLO-RV}

The real--virtual contribution to the differential decay rate reads
\beq
\dgam{RV}{3} = 
	\frac{1}{2m_H} \PS{3}(m_H^2)\, 
	2\Re\braket{\bbar g}{(0)}{}{(1)}\,,
\label{eq:GRVdiff}
\eeq
which contains explicit $\ep$-poles coming from the one-loop matrix element and furthermore it is divergent in phase space regions where the gluon becomes 
unresolved. 
The explicit poles are cancelled by the integral of 
the single unresolved subtraction term in the double real emission contribution 
to the full NNLO decay rate,
\beq
\int_1 \dgama{RR}{1}{4} = 
	\dgam{R}{3} \otimes \bI_1^{(0)}(p_1,p_2,p_3;\ep),
\label{eq:INTGRRdiffA1}
\eeq
where the real emission differential decay rate is $\dgam{R}{3}$ 
is given by \eqn{eq:GRdiff}, while the insertion operator 
$\bI_1^{(0)}(p_1,p_2,p_3;\ep)$ is given by \eqn{eq:I10}. 
As there are only three partons in the final state, the colour 
connections that appear in the generic case in \eqn{eq:I10} factorize
completely,
\beq
\bT_1\bT_2 = \frac{\CA-2\CF}{2}
\qquad\mbox{and}\qquad
\bT_1\bT_3 = \bT_2\bT_2 = -\frac{\CA}{2}\,.
\eeq
Thus,
\beq
\bsp
\bI_1^{(0)}(p_1,p_2,p_3;\ep) &= 
	\frac{\as}{2\pi}\frac{\Sep}{\Fep}
	\bigg(\frac{\mu^2}{m_H^2}\bigg)^\ep
	\bigg\{
	\CF\bigg[
		\IcC{1,q}{(0)}(y_{1Q};\ep)
		+ \IcC{1,q}{(0)}(y_{2Q};\ep)
		- 2\IcS{1}{(0),(1,2)}(Y_{12,Q};\ep) 
		\bigg]
\\&
	+\CA\bigg[
		\IcC{1,g}{(0)}(y_{3Q};\ep)
 		+ \IcS{1}{(0),(1,2)}(Y_{12,Q};\ep) 
		- \IcS{1}{(0),(1,3)}(Y_{13,Q};\ep) 
		- \IcS{1}{(0),(2,3)}(Y_{23,Q};\ep)
		\bigg]
	\bigg\}\,.
\esp
\label{eq:I103j}
\eeq
Using the expressions in \appx{appx:I103j}, it is straightforward to check that
\beq
\bsp
\bI_1^{(0)}(p_1,p_2,p_3;\ep) = 
	\frac{\as}{2\pi}\frac{\Sep}{\Fep}
	\bigg(\frac{\mu^2}{m_H^2}\bigg)^\ep
	\bigg\{&
	\frac{2\CF + \CA}{\ep^2}	
	+ \frac{1}{\ep}
		\bigg[
		(\CA-2\CF)\Log{y_{12}} 
		- \CA (\Log{y_{13}} + \Log{y_{23}})
\\&
		+\frac{11}{6} \CA + 3\CF - \frac{2}{3}\Nf \TR
		\bigg]
	+ \Oe{0}
	\bigg\}\,,
\esp
\label{eq:I103jexp}
\eeq
hence the combination
\beq
\dgam{RV}{3} + \int_1 \dgama{RR}{1}{4}
\label{eq:RV3plusI1RR14}
\eeq
is finite in $\ep$. 

Nevertheless, \eqn{eq:RV3plusI1RR14} is still singular in the single unresolved 
regions of phase space and requires regularization. We achieve this by subtracting 
two suitably defined approximate decay rates, 
$\dgama{RV}{1}{3}$ and $\left(\int_1\dgama{RR}{1}{3}\right)^{\rm{A}_1}$.
First, we consider
\beq
\dgama{RV}{1}{3} = 
	\frac{1}{2m_H} \PS{3}(m_H^2)\, 
	\cA_1 2\Re\braket{\bbar g}{(0)}{}{(1)}{}\,,
\label{eq:GRVA1diff}
\eeq
which matches the kinematic singularity structure of $\dgam{RV}{3}$.
The general definition of the real--virtual counterterm is~\cite{Somogyi:2006db},
\beq
\bsp
\cA_1 2\Re \braket{m+1}{(0)}{}{(1)} &=
\sum_{r=1}^{m+1} \Bigg[
\sum_{\substack{i=1 \\ i\ne r}}^{m+1} \frac{1}{2} \cC{ir}{(0,1)}
+ \Bigg(\cS{r}{(0,1)} 
- \sum_{\substack{i=1 \\ i\ne r}}^{m+1} \cCS{ir}{r}{(0,1)}\Bigg) \Bigg]
\\ &
+ \sum_{r=1}^{m+1} \Bigg[
\sum_{\substack{i=1 \\ i\ne r}}^{m+1} \frac{1}{2} \cC{ir}{(1,0)}
+ \Bigg(\cS{r}{(1,0)} 
- \sum_{\substack{i=1 \\ i\ne r}}^{m+1} \cCS{ir}{r}{(1,0)}\Bigg) 
\Bigg]\,.
\label{eq:A11M0M1}
\esp
\eeq
The basic structure of this subtraction in terms of unresolved limits is the 
same as the tree level single unresolved counterterm in \eqn{eq:A1def}. However, 
in accordance with the form of infrared factorization of one-loop QCD matrix elements 
\cite{Bern:1997sc,Kosower:1999xi,Kosower:1999rx,Bern:1999ry}, in \eqn{eq:A11M0M1} 
we have terms with tree level collinear or soft functions multiplying (in colour 
or spin space) one-loop matrix elements (those with the $(0,1)$ superscript),
as well as terms with one-loop collinear or soft functions multiplying tree level 
matrix elements (denoted with the $(1,0)$ superscript). The precise 
definitions of the functions appearing in \eqn{eq:A11M0M1} are given in 
\refr{Somogyi:2006db}.

Then we consider the counterterm,
\beq
\Big(\int_1\dgama{RR}{1}{4}\Big)\strut^{\rm{A}_1} =
	\frac{1}{2m_H} \PS{3}(m_H^2)\, \cA_1 
	\Big(\SME{\bbar g}{(0)}{} \otimes \bI_1^{(0)}\Big)\,,
\label{eq:GRRdiffA1A1}
\eeq
which matches the kinematic singularity structure of
$\int_1\dgama{RR}{1}{4}$.
In general, the counterterm is given by~\cite{Somogyi:2006db},
\beq
\bsp
 \cA_1  \Big(\SME{m+1}{(0)}{} \otimes \bI_1^{(0)} \Big) &=
\sum_{r=1}^{m+1} \Bigg[
\sum_{\substack{i=1 \\ i\ne r}}^{m+1} \frac{1}{2} \cC{ir}{(0,0 \otimes I)}
+ \Bigg(\cS{r}{(0,0 \otimes I)}
- \sum_{\substack{i=1 \\ i\ne r}}^{m+1} \cCS{ir}{r}{(0,0 \otimes I)}\Bigg) 
\Bigg]
\\&
+ \sum_{r=1}^{m+1} \Bigg[
\sum_{\substack{i=1 \\ i\ne r}}^{m+1} \frac{1}{2} \cC{ir}{R\times(0,0)}
+ \Bigg(\cS{r}{R\times(0,0)}
- \sum_{\substack{i=1 \\ i\ne r}}^{m+1} \cCS{ir}{r}{R\times(0,0)}\Bigg) 
\Bigg]\,.
\label{eq:A10M0I}
\esp
\eeq
The organization of this subtraction in terms of unresolved limits is again 
identical to the tree level single unresolved counterterm in \eqn{eq:A1def}. 
However, for each limit, we have two types of terms, labeled by the different 
superscripts. The reason is as follows. This counterterm is built from the 
infrared factorization formulae for the product of a QCD squared matrix element 
times the $\bI_1^{(0)}$ insertion operator of \eqn{eq:I10}. It turns out that 
these factorization formulae are sums of two pieces. Both of these involve the 
product of a tree level collinear or soft function times a tree level matrix 
element, but one piece is further multiplied by the $\bI_1^{(0)}$ insertion 
operator appropriate to the reduced matrix element, while the other is 
multiplied with a well-defined remainder function $R$~\cite{Somogyi:2006db}. 
Hence the superscripts on the various terms in \eqn{eq:A10M0I}.

It can be shown that the combination
\beq
\dgam{NNLO}{3}\equiv
	\bigg[\dgam{RV}{3} + \int_1 \dgama{RR}{1}{4}\bigg] J_3 
	-\bigg[\dgama{RV}{1}{3} +\Big(\int_1\dgama{RR}{1}{4}\Big)\strut^{\rm{A}_1}\bigg] J_2
\label{eq:dgamNNLO_3}
\eeq
is both free of $\ep$-poles and integrable in all kinematically singular
limits~\cite{Somogyi:2006db}. Thus, the regularized real--virtual contribution 
to the decay rate
\beq
\tgam{NNLO}_3[J] = 
	\int_3 
	\big[\dgam{NNLO}{3}\big]_{\ep=0}
\label{eq:tgammaNNLO_3}
\eeq
is finite and can be computed numerically in four dimensions for any 
infrared-safe observable.
For the total cross section ($J=1$) at $\mu=m_H$ ($L=0$) we find,
\beq
\tgam{NNLO}_3[J=1] =
     \tgam{LO}\,\left(\frac{\as}{\pi}\right)^2
     69.35(1)\,.
\label{eq:tgamNNLO_3}
\eeq
As for the double real emission contribution, the numerical result of the 
Monte Carlo program in \eqn{eq:tgamNNLO_3} has been reproduced by 
integrating the real--virtual matrix element and the subtraction 
terms separately in $d$ dimensions and summing the contributions. 

%
%

\subsection{Double virtual contribution}
\label{ssec:NNLO-VV}

The double virtual contribution to the differential decay rate reads
\beq
\dgam{VV}{2} = 
	\frac{1}{2m_H} \PS{2}(m_H^2)\, 
	\left[ 2\Re\braket{\bbar}{(0)}{}{(2)} 
	+ \SME{\bbar}{(1)}{} \right]
	\,,
\label{eq:GVVdiff}
\eeq
which contains explicit $\ep$-poles coming from the two-loop matrix element
and the square of the one-loop matrix element:
\beq
\bsp
\dgam{VV}{2} &= \dgam{B}{} \left(\frac{\as}{2\pi}\frac{\Sep}{\Fep}\right)^2
	\bigg\{
	\frac{2\CF^2}{\ep^4}
	+ \bigg[
	\frac{11 \CA \CF}{4} 
	+ (6 + 4 L) \CF^2
	- \Nf \TR \CF
	\bigg]\frac{1}{\ep^3} 
\\ &\quad
	+ \bigg[
	\bigg(
	\frac{8}{9}
	+ \frac{\pi^2}{12}
	+ \frac{11}{6} L
	\bigg)\CA \CF
	+\bigg(
	\frac{17}{2}
	-2\pi^2
	+6L
	+4L^2
	\bigg)\CF^2
	-\bigg(
	\frac{4}{9}
	+ \frac{2}{3}L
	\bigg) \Nf \TR \CF
	\bigg]\frac{1}{\ep^2}	
\\ &\quad
	+\bigg[
	\bigg(
	-\frac{961}{216} 
	+ \frac{13 \ze{3}}{2}
	-\frac{1}{18}(67 - 3\pi^2)L
	\bigg)\CA \CF
\\ &\quad\quad	
	+\bigg(
	\frac{109}{8}
	-2\pi^2
	-14\ze{3}
	+4(2-\pi^2)L
	+3L^2
	+\frac{8}{3}L^3
	\bigg)\CF^2
\\ &\quad\quad	
	+\bigg(
	\frac{65}{54} 
	+\frac{10}{9}L
	\bigg)\Nf \TR \CF
	\bigg]\frac{1}{\ep} 
	+ \Oe{0} \bigg\}\,.
\esp
\label{eq:GVVdiff-exp}
\eeq
The finite part of $\dgam{VV}{2}$ is also known exactly
\cite{Ravindran:2006cg} which we recall in \appx{appx:MEs} 
(see \eqns{eq:bb2loop}{eq:bb11loop}). In order to regulate these poles, 
we add the integrals of the counterterms which have been subtracted in 
\sects{ssec:NNLO-RR}{ssec:NNLO-RV}. The KLN theorem then ensures that, 
provided the physical observable we are to compute is infrared-safe
and our subtraction scheme is internally consistent,
the ensuing result will be free of infrared divergences.
It is our task in this section to verify that this is indeed the case.

Let us begin with the integral of the double unresolved subtraction 
term, \eqn{eq:GRRA2diff}, which can be written as,
\beq
\int_2 \dgama{RR}{2}{m+2} = 
	\dgam{B}{m} \otimes \bI_2^{(0)}(\{p\}_m;\ep)\,,
\label{eq:INTGRRA2diff}
\eeq
where the insertion operator has five contributions according to 
the possible colour structures,
\beq 
\bsp 
\bI_2^{(0)}(\{p\};\ep) &=  
	\left[\frac{\as}{2\pi} \frac{\Sep}{\Fep}
     \left(\frac{\mu^2}{Q^2}\right)^{\ep}\right]^2 
	\Bigg\{ \sum_{i=1}^m \Bigg[ 
	\IcC{2,i}{(0)}(y_{iQ};\ep) \, \bT_i^2 
	+ \sum_{\substack{j=1 \\ j\ne i}}^m 
	\IcC{2,i j}{(0)}(y_{iQ},y_{jQ},Y_{ij,Q};\ep)\, \bT_j^2 \Bigg] 
	\bT_i^2 
\\ & \quad 
	+ \sum_{\substack{j,l=1 \\ l\ne j}}^m \Bigg[ 
	\IcS{2}{(0),(j,l)}(Y_{jl,Q};\ep)\, \CA\,  
	+ \sum_{i=1}^m \IcSCS{2,i}{(0),(j,l)}(y_{iQ},Y_{ij,Q},Y_{il,Q},Y_{jl,Q};\ep)\, 
	\bT_i^2\, \Bigg] \bT_{j}\bT_{l} 
\\ & \quad
	+ \sum_{\substack{i,k=1, \\ k\ne i}}^m\sum_{\substack{j,l=1, \\ l\ne j}}^m 
	\IcS{2}{(0),(i,k)(j,l)}
	(Y_{ik,Q},Y_{ij,Q},Y_{il,Q},Y_{jk,Q},Y_{kl,Q},Y_{jl,Q};\ep) 
	\{ \bT_{i}\bT_{k},\bT_{j}\bT_{l} \} 
	\Bigg\} \,.
\label{eq:I2} 
\esp 
\eeq 
The kinematic functions in \eqn{eq:I2} have been defined and computed
as expansions in $\ep$ in \refrs{DelDuca:2013kw,Somogyi:2013yk}.  Again,
there is no one-to-one correspondence between the unintegrated double
unresolved subtraction terms in \eqn{eq:RR_A2} and the kinematic
functions that appear in \eqn{eq:I2}. The latter are obtained from the
former after integration over unresolved momenta and summation over
unobserved colours and flavours. This remark applies to the rest of
the insertion operators to be discussed below.

For $H \to \bbar$, the colour connections that appear in \eqn{eq:I2}
are simply given by \eqn{eq:connect2}, and the kinematic variables
simplify as in \eqn{eq:Bornkin}. Furthermore, when evaluating
\eqn{eq:I2} the coincidence of certain summation indices is allowed. In
particular, $i$ in the second line need not be distinct from $j$ and
$l$, while in the last line we only require that $i$ and $k$ as well as
$j$ and $l$ are different, with no further restrictions, as shown in
the formula. As a result, some indices of kinematic functions coincide
once we explicitly write out \eqn{eq:I2}. Specifically, since in our
case there are only two hard partons in the final state, only
$\IcSCS{2,i}{(0),(i,l)}$ and $\IcS{2}{(0),(i,k),(i,k)}$ appear, while
the more general functions $\IcSCS{2,i}{(0),(j,l)}$ or
$\IcS{2}{(0),(i,k),(j,l)}$ are absent from the sum, as those require at
least three hard partons if all indices are different. In such cases we
also simplify the list of arguments of the functions so that we do not
display arguments that are the same or identically zero. For instance,
in $\IcSCS{2,i}{(0),(j,l)}$ if $i=j$, then $Y_{ij,Q} = 0$ and $Y_{il,Q}
= Y_{jl,Q}$. Hence, $\IcSCS{2,i}{(0),(i,l)}$ is a function of $y_{iQ}$
and $Y_{il,Q}$ only.  Similarly $\IcS{2}{(0),(i,k),(i,k)}$ depends just
on the variable $Y_{ik,Q}$.  Then, we obtain the
$\bI_2^{(0)}(p_1,p_2;\ep)$ operator,
\beq
\bsp
\bI_2^{(0)}(p_1,p_2;\ep)   =
	\left[\frac{\as}{2\pi}  \frac{\Sep}{\Fep}
	\left(\frac{\mu^2}{m_H^2}\right)^\ep\right]^2
	&\bigg\{
	2 \CF^2\bigg[\rC^{(0)}_{2,q}(1;\ep) + \rC^{(0)}_{2,q q}(1,1,1;\ep)
	- 2 \rSCS^{(0),(1,2)}_{2,q}(1,1;\ep) 
\\&
	+ 4 \rS^{(0),(1,2)(1,2)}_{2}(1;\ep)\bigg]
	- 2 \CF \CA \rS^{(0),(1,2)}_{2}(1;\ep)\bigg\}\,,
\esp
\label{eq:I22jet}
\eeq
whose $\ep$-expansion is
\beq
\bsp
\bI_2^{(0)}(p_1,p_2;\ep) &= 
        \left[ \frac{\as}{2\pi} \frac{\Sep}{\Fep}
        \bigg(\frac{\mu^2}{m_H^2}\bigg)^\ep \right]^2  
	\bigg\{
        \left( 
        \frac{\CA\CF}2 
        + 2\CF^2
        \right) \frac1{\ep^4}
        + \left( 
        \frac{29\CA\CF}{12} 
        + 6\CF^2      
        - \frac{\Nf\TR\CF}3 
        \right) \frac1{\ep^3} 
\\& \quad 
		+ \left[ \left( 
       	\frac{68}9 
		- \frac{7\pi^2}{12} 
		\right) \CA\CF 
        + \left( 
        \frac{170}9 
        - \frac{8\pi^2}3 
        \right) \CF^2 
        - \frac{14\Nf\TR\CF}9 
        \right] \frac1{\ep^2} 
\\
        & \quad +
        \bigg[ \left( 
        -\frac{301}{216}
        - \frac{37\pi^2}{12}
        + \frac{\ze{3}}{2}  
        \right) \CA\CF 
        + \left( 
        \frac{6149}{216}
        - \frac{47\pi^2}{18}
        -70\ze{3}  
        \right) \CF^2 
\\
        & \quad\quad + 
        \left(-\frac{97}{18} 
        +\frac{5\pi^2}9          
        \right) \Nf\TR\CF  
        \bigg] \frac1{\ep} 
\\ 
        & \quad - 227.559 \CA\CF - 236.532 \CF^2 + 30.9273 \Nf\TR\CF
        + \Oe{} \bigg\} \,.
\esp
\label{eq:I22jet-exp}
\eeq
The coefficients of the poles are all given in terms of rational numbers and 
known transcendental constants.

Next, we consider the integral of the iterated single unresolved subtraction 
term, \eqn{eq:GRRA12diff}, which can be written as,
\beq
\int_2 \dgama{RR}{12}{m+2} =
	\dgam{B}{m} \otimes \bI_{12}^{(0)}(\{p\}_m;\ep)\,,
\label{eq:INTGRRA12diff}
\eeq
where the insertion operator in general has the same structure in colour and 
flavour space as $\bI_2^{(0)}$ in \eqn{eq:I2},
\beq
\bsp
\bI_{12}^{(0)}(\{p\};\ep) &= 
\left[\frac{\as}{2\pi}  \frac{\Sep}{\Fep}
\left(\frac{\mu^2}{Q^2}\right)^\ep\right]^2
\bigg\{
\sum_{i=1}^m
	\bigg[\IcC{12,\fla{i}}{(0)}(y_{iQ};\ep)\,\bTsq{\fla{i}} 
	+ \sum_{\substack{k=1 \\ k\ne i}}^m 
	\IcC{12,\fla{i}\fla{k}}{(0)}(y_{iQ},y_{jQ},Y_{ij,Q};\ep)
	\,\bTsq{\fla{k}} \bigg] 
	\bTsq{\fla{i}}
\\& \quad
+\sum_{\substack{j,l=1 \\ l\ne j}}^m
	\bigg[\IcS{12}{(0),(j,l)}(Y_{jl,Q};\ep) \CA 
	+ \sum_{i=1}^m \IcSCS{12,\fla{i}}{(0),(j,l)}(y_{iQ},Y_{ij,Q},Y_{il,Q},Y_{jl,Q};\ep)
	 \bTsq{\fla{i}} \bigg] \bT_j \bT_l
\\& \quad
+\sum_{\substack{i,k=1 \\ k\ne i}}^m \sum_{\substack{j,l=1 \\ l\ne j}}^m
	\IcS{12}{(0),(i,k)(j,l)}
	(Y_{ik,Q},Y_{ij,Q},Y_{il,Q},Y_{jk,Q},Y_{kl,Q},Y_{jl,Q};\ep) 
	\{\bT_i \bT_k , \bT_j \bT_l\}
\bigg\}\,.
\label{eq:I12}
\esp
\eeq
The kinematic functions in \eqn{eq:I12} have been defined
and computed as expansions in $\ep$ in \refr{Bolzoni:2010bt}.
The discussion below \eqn{eq:I2} applies to \eqn{eq:I12} as well, hence,
using \eqns{eq:connect2}{eq:Bornkin}, we obtain the $\bI_{12}^{(0)}(p_1,p_2;\ep)$ operator,
\beq
\bsp
\bI_{12}^{(0)}(p_1,p_2;\ep)  =
	\left[\frac{\as}{2\pi}  \frac{\Sep}{\Fep}
	\left(\frac{\mu^2}{m_H^2}\right)^\ep\right]^2
	&\bigg\{
	2 \CF^2\bigg[\rC^{(0)}_{12,q}(1;\ep) + \rC^{(0)}_{12,q q}(1,1,1;\ep)
	- 2 \rSCS^{(0),(1,2)}_{12,q}(1,1;\ep) 
\\&
	+ 4 \rS^{(0),(1,2)(1,2)}_{12}(1;\ep)\bigg]
	- 2 \CF \CA \rS^{(0),(1,2)}_{12}(1;\ep)\bigg\}\,,
\label{eq:I122jet}
\esp
\eeq
whose $\ep$-expansion is
\beq
\bsp
\bI_{12}^{(0)}(p_1,p_2;\ep) &= 
        \left[ \frac{\as}{2\pi} \frac{\Sep}{\Fep}
        \bigg(\frac{\mu^2}{m_H^2}\bigg)^\ep \right]^2
	\bigg\{
        \frac{4\CF^2}{\ep^4} 
        + \left( 
        - \frac{\CA\CF}3 
        + 12\,\CF^2 
        - \frac{2\Nf\TR\CF}3 
        \right) \frac1{\ep^3} 
\\
        & \quad + 
        \left[ \left( 
        - \frac{155}{18} 
        + \pi^2 
        \right) \CA\CF 
        + \left( 
        \frac{788}{25} 
        - \frac{16\pi^2}3 
        \right) \CF^2 
        - \frac{31\Nf\TR\CF}9 
        \right] \frac1{\ep^2} 
\\ 
        & \quad +
        \bigg[ \left(  
        - \frac{5911}{54} 
        + \frac{101\Log{2}}9
        + \frac{49\pi^2}6 
        + 42 \ze{3} 
        \right) \CA\CF 
        - \left( 
        \frac{116497}{4500} 
        + \frac{296\pi^2}{45} 
        + 104 \ze{3} 
        \right) \CF^2
\\
        & \quad\quad    
        + \left( 
        \frac{71}{36} 
        - \frac{202\Log{2}}9 
        + \frac{8\pi^2}9 
        \right) \Nf\TR\CF
        \bigg] \frac1{\ep} 
\\
        & \quad + 
        215.508 \CA\CF 
        - 717.881 \CF^2 
        + 22.1494 \Nf\TR\CF
        + \Oe{} \bigg\}  \,.
\esp
\label{eq:I122jet-exp}
\eeq
As in the case of $\bI_{2}^{(0)}$, the coefficients of the poles are all 
given in terms of rational numbers and known transcendental constants.

Turning to the integral of the real--virtual single unresolved subtraction term,
\eqn{eq:GRVA1diff}, we find~\cite{Somogyi:2008fc}
\beq
\int_1 \dgama{RV}{1}{m+1} =
  \dgam{V}{m} \otimes \bI_{1}^{(0)}(\{p\}_m;\ep)
+ \dgam{B}{m} \otimes \bI_{1}^{(1)}(\{p\}_m;\ep)
\,,
\label{eq:I1dsigRVA1}
\eeq
where the insertion operator $\bI_{1}^{(0)}$ is given in \eqn{eq:I10},
expanded to sufficiently high order in \eqn{eq:I10-2jetexp} to obtain the 
first term on the right-hand side in \eqn{eq:I1dsigRVA1} to $\Oe{}$, while 
the $\bI_{1}^{(1)}$ operator in general reads
\beq
\bI_{1}^{(1)}(\{p\}_m;\ep) =
\bI_{1}^{(1),B}(\{p\}_m;\ep)
-\frac{\as}{2\pi}\frac{\beta_0}{2\ep}
\bI_{1}^{(0)}(\{p\}_m;\ep) \,.
\eeq
The unrenormalized operator $\bI_{1}^{(1),B}$ has the following 
structure in colour and flavour space,
\beq
\bsp
\bI_{1}^{(1),B}(\{p\}_m;\ep) &=
	\left[\frac{\as}{2\pi}  \frac{\Sep}{\Fep}
	\left(\frac{\mu^2}{Q^2}\right)^\ep\right]^2
	\sum_{i=1}^m\bigg[
	\IcC{1,\fla{i}}{(1),B}(y_{iQ};\ep)\,\CA \bTsq{i}
	+\sum_{\substack{k=1\\ k\ne i}}^m \IcS{1}{(1),(i,k),B}(\Y{i}{k};\ep)\,\CA \bT_{i}\bT_{k}
\\ & \quad
	+\sum_{\substack{k=1\\ k\ne i}}^m \sum_{\substack{l=1\\ l\ne i,k}}^m
	\IcS{1}{(1),(i,k,l),B}(\Y{i}{k},\Y{i}{l},\Y{k}{l};\ep)
	\sum_{a,b,c}f_{abc}T_i^a T_k^b T_l^c
	\bigg]\,.
\esp
\label{eq:I11-bare}
\eeq
The bare kinematic functions in \eqn{eq:I11-bare} have been defined 
and computed as expansions in $\ep$ in \refr{Somogyi:2008fc}.
Using \eqns{eq:connect2}{eq:Bornkin}, the unrenormalized 
$\bI_{1}^{(1),B}(p_1,p_2;\ep)$ operator becomes
\beq
\bI_{1}^{(1),B}(p_1,p_2;\ep) = 
	\left[\frac{\as}{2\pi}  \frac{\Sep}{\Fep}
	\left(\frac{\mu^2}{m_H^2}\right)^\ep\right]^2
	\bigg\{2\CA\CF\bigg[
	\IcC{1,q}{(1),B}(1;\ep) 
	-\IcS{1}{(1),(1,2),B}(1;\ep)\bigg]
	\bigg\}\,.
\label{eq:I11-2jet}
\eeq
The term involving triple colour correlations on the second line of 
\eqn{eq:I11-bare} does not contribute, the triple sum over $i$, $k$ and $l$ 
being empty because we cannot form a triplet of distinct indices.
The $\ep$-expansion of the bare insertion operator $\bI_{1}^{(1),B}$ 
reads
\beq
\bsp
\bI_1^{(1),B}(p_1,p_2;\ep) &= 
	\left[\frac{\as}{2\pi}  \frac{\Sep}{\Fep}
	\left(\frac{\mu^2}{m_H^2}\right)^\ep\right]^2
	\bigg\{
        -\frac{\CA\CF}{2\ep^4} 
        - \frac{3\CA\CF}{2\ep^3} 
\\& \quad 
        + 
        \bigg[ \bigg( 
        \frac{83}{450} 
        + \frac{\pi^2}{2} 
        \bigg) \CA\CF 
        + \bigg( 
        -\frac{5}{2} 
        + \frac{2\pi^2}{3} 
        \bigg) \CF^2 
        \bigg] \frac1{\ep^2} 
\\
        & \quad + 
        \bigg[ \bigg( 
        \frac{580571}{6750} 
        - \frac{43\pi^2}{30} 
        -15 \ze{3}
        \bigg) \CA\CF 
        + \bigg( 
        \frac{661}{50} 
        - \frac{13184 \Log{2}}{225}
        + \frac{71\pi^2}{45} 
        + 38 \ze{3}
        \bigg) \CF^2 
        \bigg] \frac1{\ep} 
\\
        & \quad + 
        292.930 \CA\CF 
        + 134.720 \CF^2 
        + \Oe{} \bigg\}  \,.
\esp
\label{eq:I112jet-exp}
\eeq
Also here we see that the pole coefficients are all given in terms of 
rational numbers and known transcendental constants.  

Finally the iterated integral of the double real single unresolved subtraction 
term, \eqn{eq:GRRdiffA1A1}, can be written as,
\beq
\int_1 \Big(\int_1\dgama{RR}{1}{m+2}\Big)\strut^{\rm{A}_1} =
\dgam{B}{m} \otimes \left[
\frac12\Big\{
\bI_{1}^{(0)}(\{p\}_m;\ep),
\bI_{1}^{(0)}(\{p\}_m;\ep)
\Big\}
+ \bI_{1,1}^{(0,0)}(\{p\}_m;\ep)
\right]
\,,
\label{eq:I1dsigRA1_A1}
\eeq
where the insertion operator $\bI_{1}^{(0)}$ is given in \eqn{eq:I10},
expanded to sufficiently high order in \eqn{eq:I10-2jetexp} to obtain
the first term on the right-hand side of \eqn{eq:I1dsigRA1_A1} to $\Oe{}$, 
while $\bI_{1,1}^{(0,0)}$ in general reads
\beq
\bI_{1,1}^{(0,0)}(\{p\}_m;\ep) =
	\left[\frac{\as}{2\pi}  \frac{\Sep}{\Fep}
	\left(\frac{\mu^2}{Q^2}\right)^\ep\right]^2
	\sum_{i=1}^m\bigg[
	\IcC{1,1,\fla{i}}{(0,0)}(y_{iQ};\ep)\,\CA\bTsq{\fl{i}}
	+\sum_{\substack{k=1\\ k\ne i}}^m
	\IcS{1,1}{(0,0),(i,k)}(\Y{i}{k};\ep)\,\CA\,\bT_{i}\bT_{k}
\bigg]\,.
\label{eq:I1100}
\eeq
The kinematic functions in \eqn{eq:I1100} have been defined and
computed as expansions in $\ep$ in \refr{Somogyi:2008fc}.
Using \eqns{eq:connect2}{eq:Bornkin}, we obtain
\beq
\bI_{1,1}^{(0,0)}(p_1,p_2;\ep) = 
	\left[\frac{\as}{2\pi}  \frac{\Sep}{\Fep}
	\left(\frac{\mu^2}{m_H^2}\right)^\ep\right]^2
	\bigg\{2\CA\CF\bigg[
	\IcC{1,1,q}{(0,0)}(1;\ep) 
	-\IcS{1,1}{(0,0),(1,2)}(1;\ep)\bigg]
	\bigg\}\,,
\label{eq:I1100-2jet}
\eeq
whose $\ep$-expansion is
\beq
\bsp
\bI_{1,1}^{(0,0)}(p_1,p_2;\ep) &= 
	\left[\frac{\as}{2\pi}  \frac{\Sep}{\Fep}
	\left(\frac{\mu^2}{m_H^2}\right)^\ep\right]^2
	\bigg\{
        - \left( \frac{\CA\CF}3 + \frac{2\Nf\TR\CF}3 \right) \frac1{\ep^3} \\
        & \quad + 
        \left[ \left( - \frac{587}{50} + \pi^2 \right) \CA\CF 
        + \left( 5 - \frac{4\pi^2}3 \right) \CF^2 - \frac{31\Nf\TR\CF}9 \right] \frac1{\ep^2} \\ 
        & \quad 
        + 
        \bigg[ \left( - \frac{622583}{3375} 
        + \frac{101\Log{2}}9 + \frac{502\pi^2}{45} + 50 \ze{3}\right) \CA\CF \\
        & \quad\quad   
        + \left( \frac{393797}{13500} + \frac{13184\Log{2}}{225} 
         - \frac{274\pi^2}{45} - 66 \ze{3} \right) \CF^2 \\
        & \quad\quad      + \left( \frac{11557}{2700} 
        - \frac{202\Log{2}}9 + \frac{8\pi^2}9 \right) \Nf\TR\CF
        \bigg] \frac1{\ep} \\
        & \quad - 15.2343 \CA\CF - 318.099 \CF^2 + 46.4407 \Nf\TR\CF
        + \Oe{} \bigg\} \,.
\esp
\label{eq:I11002jet-exp}
\eeq
All the pole coefficients are again given in terms 
of rational numbers and known transcendental constants.

Using eqs.~(\ref{eq:GVVdiff-exp}),  (\ref{eq:I22jet-exp}), (\ref{eq:I122jet-exp}), 
(\ref{eq:I112jet-exp}) and (\ref{eq:I11002jet-exp}), it is straightforward to 
check that the regularized double virtual contribution
\beq
\dgam{NNLO}{2} \equiv
	\bigg\{
	\dgam{VV}{2} 
	+ \int_2 \bigg[\dgama{RR}{2}{4} - \dgama{RR}{12}{4}\bigg]
	+ \int_1 \bigg[\dgama{RV}{1}{3}
		+ \bigg(\int_1\dgama{RR}{1}{4}\bigg)^{\rm{A}_1}\bigg]\bigg\} J_2
\label{eq:dgamNNLO_2}
\eeq
is free of $\ep$-poles. Hence, the regularized double virtual contribution 
to the decay rate
\beq
\tgam{NNLO}_2[J] = \int_2 \big[\dgam{NNLO}{2}\big]_{\ep=0}
\eeq
is finite for any infrared-safe observable and can be computed numerically in 
four dimensions.
For the total cross section ($J=1$) at $\mu=m_H$ ($L=0$) we find,
\beq
\tgam{NNLO}_2[J=1] =
     -\tgam{LO}\,\left(\frac{\as}{\pi}\right)^2
     41.25(1)\,.
\label{eq:tgamNNLO_2}
\eeq
We note that the error estimate of the above result comes entirely from the 
uncertainty associated with the numerical computation of the finite parts 
of the insertion operators. The statistical uncertainty of the Monte Carlo 
integration over the two-parton phase space is completely negligible.

Finally, summing eqs.~(\ref{eq:tgamNNLO_4}), (\ref{eq:tgamNNLO_3}) 
and (\ref{eq:tgamNNLO_2}), we obtain
\beq
\tgam{NNLO}[J=1] = \tgam{NNLO}_4[1] + \tgam{NNLO}_3[1] + \tgam{NNLO}_2[1] =
     \tgam{LO}\,\left(\frac{\as}{\pi}\right)^2
     29.15(2)\,,
\label{eq:tgamNNLO}
\eeq
to be compared with the know analytic result
\beq
\tgam{NNLO}[J=1] = 
     \tgam{LO}\,\left(\frac{\as}{\pi}\right)^2
     29.146714\ldots\,.
\eeq


\section{Inclusive and differential results}
\label{sec:numres}

In this section, we show that using the fully differential
two-, three- and four-parton contributions of eqs.~(\ref{eq:dgamLO_2}), (\ref{eq:dgamNLO_3}), (\ref{eq:dgamNLO_2}),
(\ref{eq:dgamNNLO_4}), (\ref{eq:dgamNNLO_3}) and (\ref{eq:dgamNNLO_2}),
we can make predictions for any infrared-safe jet cross section with
jet functions $J_n$ ($n=2$, 3 and 4) defined in $d=4$ dimensions.

\begin{figure}
\begin{center}
\includegraphics[width=0.78\textwidth]{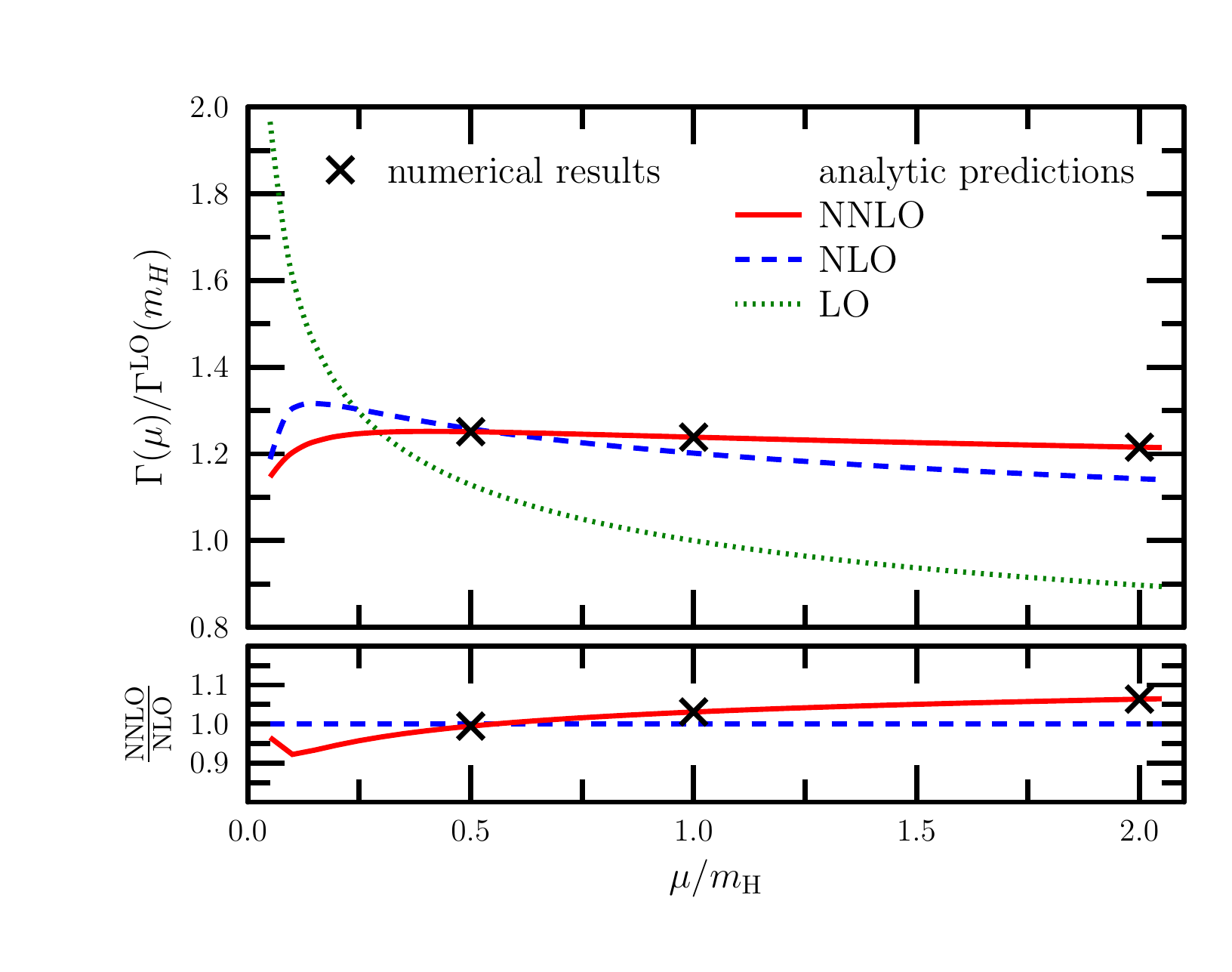}
\caption{\label{fig:tgam-mudep} 
Scale dependence of the inclusive decay rate at LO, NLO and NNLO accuracy. 
The estimated uncertainty on the numerical results is too small to be 
appreciated. 
}
\end{center}
\end{figure}

The inclusive decay rate is obtained by setting $J=1$ and is given by the sum of 
the leading order width (\ref{eq:GB}) and the NLO (\ref{eq:tgamNLO})
and NNLO (\ref{eq:tgamNNLO}) corrections. At $\mu = m_H$ we obtain
\beq
\Gamma_{\rm NNLO} = \tgam{LO}\,
\left[1
+ \frac{\as}{\pi}\,\frac{17}{3}
+ \left(\frac{\as}{\pi}\right)^2 29.15(2)
\right]
\,,
\eeq
in agreement with the known analytic 
prediction~\cite{Gorishnii:1990zu,Gorishnii:1991zr,Baikov:2005rw}.
In \fig{fig:tgam-mudep},
we compute the inclusive decay rate at $\mu = m_H/2$ and $\mu = 2 m_H$
and compare it to the known analytic result for the scale dependence, 
finding excellent agreement.

\begin{figure}[h]
\begin{center}
\includegraphics[width=0.78\textwidth]{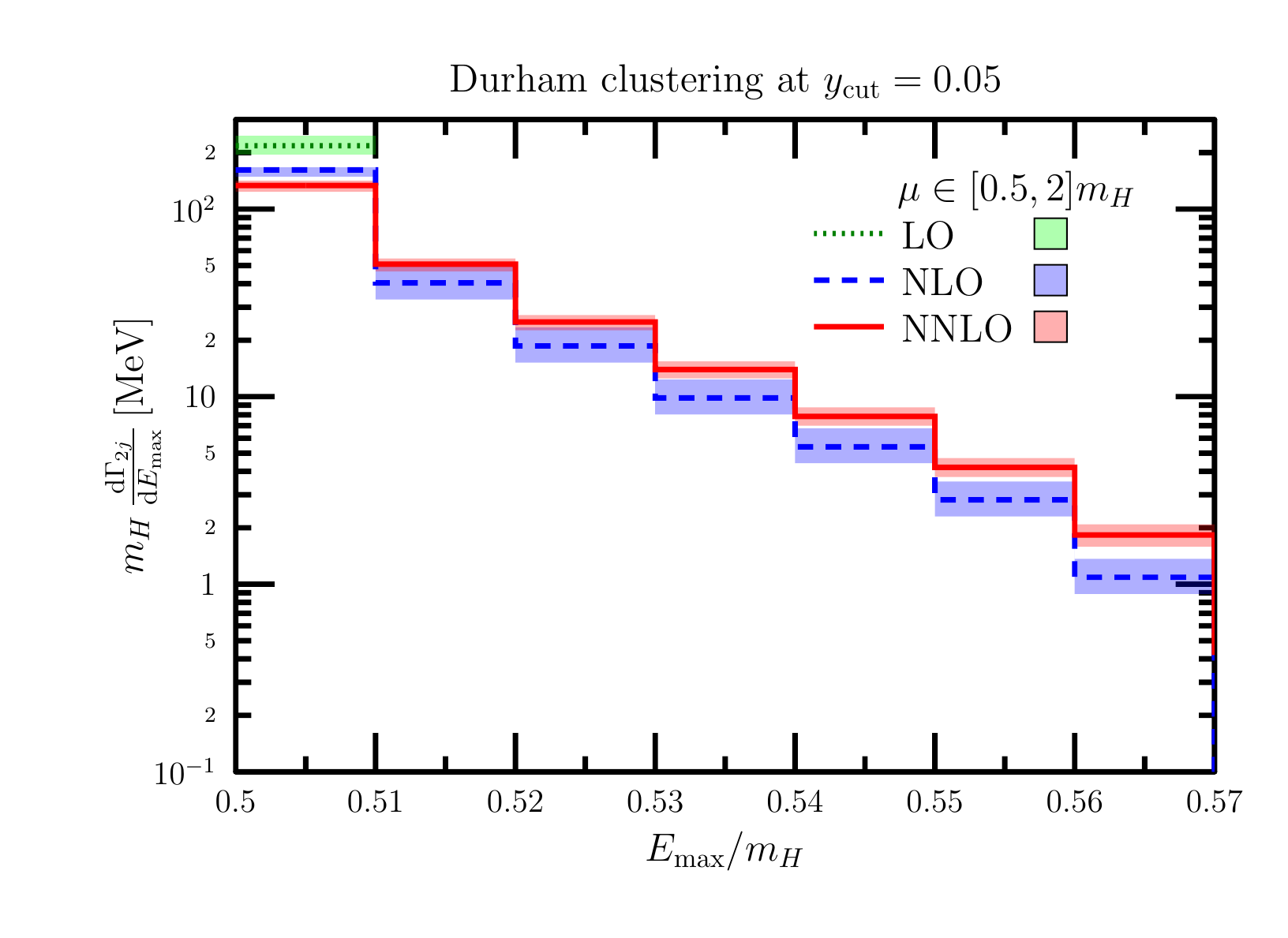}
\includegraphics[width=0.78\textwidth]{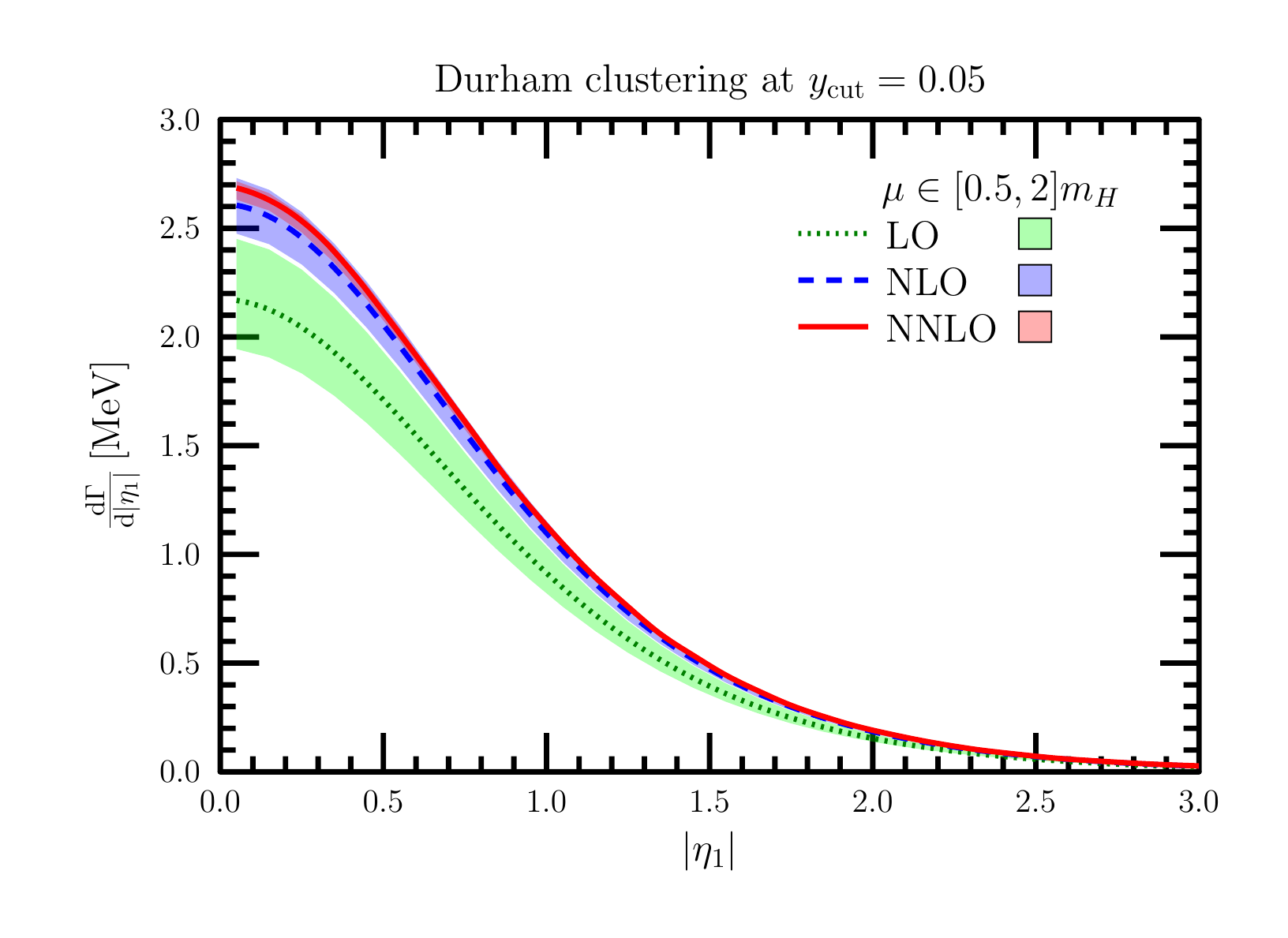}
\caption{\label{fig:durham-plots} 
The plots show the normalized distribution of the leading jet energy 
$E_{\max}$ (top) and the distribution of the absolute value of the 
pseudorapidity $|\eta_1|$ of the highest energy jet (bottom) at LO, NLO 
and NNLO accuracy. The bands show the dependence on the renormalization 
scale corresponding to the range $\mu \in [m_H/2, 2 m_H]$. Jets have 
been clustered using the Durham algorithm, the resolution parameter for 
jet clustering was set to $\yc = 0.05$.}
\end{center}
\end{figure}

To illustrate the impact of NNLO QCD corrections on differential 
distributions, we apply the Durham jet algorithm \cite{Catani:1992ua} 
with resolution parameter $\yc=0.05$ to cluster final state partons 
and order the resulting jets in energy. In the top panel of 
\fig{fig:durham-plots} we show the energy distribution of the leading 
jet in the rest frame of the decaying Higgs boson for two-jet events. 
In \refr{Anastasiou:2011qx} the same distribution was computed for jets 
clustered according to the JADE algorithm with $\yc=0.1$. We have repeated 
that calculation and found excellent agreement with the published results.
However, for two-parton kinematics the energy of the leading jet is just 
$E_{\max} = m_H/2$, so at leading order the leading jet energy distribution 
is a delta function. Furthermore, double unresolved subtractions for four 
parton matrix elements, as well as single unresolved subtractions for three 
parton matrix elements also contribute to this distribution only at 
$E_{\max} = m_H/2$. 
Then, to show the subtraction method at work on an observable that has a 
non-trivial distribution already at leading order, we consider the absolute 
value of the pseudorapidity of the leading jet, $|\eta_1|$, with respect to an 
arbitrary axis. The effect of higher order corrections on this distribution 
is shown on the bottom panel of \fig{fig:durham-plots}. In this last illustrative 
example we note that going from the leading order to NNLO, the uncertainty 
bands shrink, and that the NNLO band falls within the NLO band, thereby showing 
the good convergence of the perturbative series.

The bands in both distributions in \fig{fig:durham-plots} correspond to the 
envelope of varying the renormalization scale in the range $\mu \in [m_H/2, 2 m_H]$.


\section{Conclusions}
\label{sec:concl}

In this paper, we have computed the fully differential decay rate 
of the SM Higgs boson into b-quarks at NNLO accuracy in $\as$,
by implementing a general subtraction scheme developed
in a series of papers for the computation of QCD jet cross sections at
NNLO accuracy~\cite{Somogyi:2005xz,Somogyi:2006cz,Somogyi:2006da,Somogyi:2006db,
Somogyi:2008fc,Aglietti:2008fe,Somogyi:2009ri,Bolzoni:2009ye,
Bolzoni:2010bt,DelDuca:2013kw,Somogyi:2013yk}. 

We have shown that our subtractions render both the double real and 
real--virtual contributions to the NNLO correction integrable in 
four dimensions. We have also presented the integrated forms of our 
subtraction terms with pole coefficients evaluated analytically, while 
the finite parts were given numerically. We confirmed that the sum of the 
double virtual contribution and the integrated subtractions is free of 
infrared singularities as required by the KLN theorem.
We have implemented our computation in a parton level Monte Carlo program 
and presented illustrative examples of differential distributions at NNLO.

The successful application of our subtraction scheme reported here opens 
the way to the computation of other, more involved processes and is also 
encouraging to further developments of the scheme to deal with initial state 
radiation. These directions of development are under way and will be the 
subject of further publications.


\section*{Acknowledgments}
We thank Franz Herzog for useful communication. 
We thank Roman Derco, Zolt\'an Sz\H or, Damiano Tommasini, Zolt\'an Tulip\'ant 
and especially \'Ad\'am Kardos for useful discussions.
This research was supported by the Hungarian Scientific Research Fund grant 
K-101482,  by the European Union and the State of Hungary, co-financed by the 
European Social Fund in the framework of T\'AMOP 4.2.4.~A/2-11-1-2012-0001 
ÔNational Excellence ProgramÕ and LHCPhenoNet network PITN-GA-2010-264564 
projects, and by the Italian Ministry of University and Research under the 
PRIN project 2010YJ2NYW and by the Istituto Nazionale di Fisica Nucleare (INFN) 
through the Iniziativa Specifica PhenoLNF. 

\newpage


\appendix


\section{Matrix elements}
\label{appx:MEs}

We present the matrix elements in the form used in our parton level 
Monte Carlo program. In particular, in our scheme we need the 
the four-parton tree level and the three-parton one-loop matrix elements 
only up to finite terms in $\ep$. Higher order terms must of course be 
included when integrating the matrix elements and subtraction terms separately 
in $d$ dimensions. When needed for our cross checks, we take these higher order 
terms directly from \refr{Anastasiou:2011qx}.

%
%

\subsection{Two partons}

For $H \to \bbar$ at tree level we have
\beq
\SME{\bbar}{(0)}{} = 
	2 \yb^2 m_H^2 \Nc\,.
\eeq
We computed the one-loop correction and obtained
\beq
\bsp
2\Re \braket{\bbar}{(0)}{}{(1)} =&
	\frac{\as}{2\pi} \frac{\Sep}{\Fep}
	\bigg(\frac{\mu^2}{m_H^2}\bigg)^\ep 
	\SME{\bbar}{(0)}{ } \CF 
	\bigg\{
		-\frac{2}{\ep^2} 
		- \frac{3}{\ep}
		-2 
		+ \pi^2 
		+ 3 L
\\&
		- \bigg(
		4 
		+ \frac{\pi^2}{4} 
		- 4\ze{3} 
		+ \frac{3}{2}L^2
		\bigg)\ep
		- \bigg(
		8 
		- \pi^2 
		+ \ze{3} 
		+ \frac{\pi^4}{60} 
		- \frac{\pi^2}{4}L 
		- \frac{1}{2}L^3
		\bigg)\ep^2 
		+ \Oe{3} \bigg\}\,.
\esp
\label{eq:bb1loop}
\eeq
We used the formula at two loops as given in \refr{Anastasiou:2011qx}:
\beq
\bsp
2\Re \braket{\bbar}{(0)}{}{(2)} =&
	\bigg[\frac{\as}{2\pi} \frac{\Sep}{\Fep}
	\bigg(\frac{\mu^2}{m_H^2}\bigg)^\ep\bigg]^2
	\SME{\bbar}{(0)}{ }
	\bigg\{
		\frac{\CF^2}{\ep^4}
		+ \bigg(
		\frac{11\CA\CF}{4} 
		+ 3\CF^2 
		- \Nf \TR \CF
		\bigg)\frac{1}{\ep^3}
\\&
		+ \bigg[
		\bigg(
		\frac{8}{9} 
		+ \frac{\pi^2}{12} 
		- \frac{11}{3} L
		\bigg)\CA \CF 
		+ \bigg(
		\frac{17}{4}
		- 2 \pi^2 
		- 3 L
		\bigg)\CF^2 
		-\bigg(
		\frac{4}{9} 
		- \frac{4}{3} L
		\bigg)\Nf \TR \CF
		\bigg]\frac{1}{\ep^2}
\\&
		+ \bigg[
		\bigg(
		- \frac{961}{216}
		+ \frac{13 \ze{3}}{2}
		- \frac{11}{2} L
		+ \frac{11}{6} L^2
		\bigg)\CA \CF 
\\&\quad
		+ \bigg(
		\frac{53}{8}
		- \frac{3 \pi^2}{4}
		- 10 \ze{3}
		- \frac{9}{2} L
		+ \frac{3}{2} L^2
		\bigg)\CF^2 
		+ \bigg(
		\frac{65}{54}
		+ 2 L
		- \frac{2}{3} L^2
		\bigg)\Nf \TR \CF
   		\bigg]\frac{1}{\ep}
\\&
		+ \bigg[	
		\bigg(
		- \frac{467}{162}
		+ \frac{733 \pi^2}{216}
		+ \frac{92\ze{3}}{9}
		- \frac{11 \pi ^4}{360}
		+ \bigg(\frac{53}{12} + \frac{55 \pi ^2}{36}\bigg) L
		+ \frac{11}{2} L^2		
		- \frac{11}{18} L^3
		\bigg)\CA \CF 
\\&\quad
   		+\bigg(
		17
		- \frac{55\pi ^2}{24}
		- 20 \ze{3}
		+ \frac{43 \pi ^4}{90}
		- \bigg(\frac{9}{4} - \frac{5 \pi^2}{4}\bigg) L
		+ \frac{9}{2} L^2
		- \frac{1}{2} L^3
		\bigg)\CF^2
\\&\quad
		+ \bigg(
		\frac{200}{81}
		- \frac{59 \pi ^2}{54}
		- \frac{4 \ze{3}}{9}
		- \bigg(\frac{1}{3} + \frac{5 \pi ^2}{9}\bigg) L
		- 2 L^2 
		+ \frac{2 L^3}{9}
		\bigg)\Nf \TR \CF
		\bigg] 
		+ \Oe{} \bigg\} 
\,.
\esp
\label{eq:bb2loop}
\eeq
We checked that the poles of this expression satisfy the 
general formula given in \refr{Catani:1998bh}, while the
finite part agrees with that in \refr{Ravindran:2006cg}.
The square of the one-loop matrix element is
\beq
\bsp
2\Re \braket{\bbar}{(1)}{}{(1)} =&
	\bigg[\frac{\as}{2\pi} \frac{\Sep}{\Fep}
	\bigg(\frac{\mu^2}{m_H^2}\bigg)^\ep\bigg]^2
	\SME{\bbar}{(0)}{ } \CF^2 
	\bigg\{
		\frac{1}{\ep^4}
		+ \frac{3}{\ep^3}
		+ \bigg(
		\frac{17}{4}
		- 3 L
		\bigg)\frac{1}{\ep^2}
\\&
		+\bigg(
		7
		- \frac{5\pi^2}{4}
		- 4 \ze{3}
		- \frac{9}{2} L
		+ \frac{3}{2} L^2
		\bigg)\frac{1}{\ep}
\\&
		+ \bigg[
		15
		+ \frac{3 \pi^2}{8}
		- 5 \ze{3}
		- \frac{\pi^4}{15}
		- \bigg(3 - \frac{5\pi^2}{4}\bigg) L
		+ \frac{9}{2} L^2
		- \frac{1}{2} L^3
		\bigg]
		+ \Oe{} \bigg\}
\,.
\esp
\label{eq:bb11loop}
\eeq

%
%

\subsection{Three partons}

For $H \to \bbar g$ at tree level we have
\beq
\SME{\bbar g}{(0)}{} = 
	8 \pi \frac{\as}{\Fep} \mu^{2\ep}
	\SME{\bbar}{(0)}{ }
	\CF
	\frac{1}{m_H^2} \bigg[
	\frac{(1-\ep) y_{23}}{y_{13}}
	+ \frac{(1-\ep) y_{13}}{y_{23}}
	+ \frac{2 y_{12}}{y_{13} y_{23}} + 2 - 2\ep\bigg]
\,.
\eeq
At one loop, we use the $\ep$-expansion of the formula from
\refr{Anastasiou:2011qx}, which we checked numerically against {\sc GoSam} 
\cite{Cullen:2011ac,Cullen:2014yla},
\beq
\bsp
2\Re \braket{\bbar g}{(0)}{}{(1)} =&
	\frac{\as}{2\pi} \frac{\Sep}{\Fep}
	\bigg(\frac{\mu^2}{m_H^2}\bigg)^\ep 
	\bigg\{
	\SME{\bbar g}{(0)}{ }
	\bigg[
		-\frac{2\CF + \CA}{\ep^2} 
		- \bigg(
		3\CF 
		+ \frac{11\CA}{6}
		-\frac{2\Nf \TR}{3}
\\&\quad
		+ (\CA - 2\CF) \Log{y_{12}} 
		- \CA (\Log{y_{13}}  + \Log{y_{23}})
		\bigg)\frac{1}{\ep}
\\&\quad
		+ (\CA - 2\CF)\bigg(
			R(y_{12},y_{13})
			+ R(y_{12},y_{23})
			+ \frac{1}{2} \Log{y_{12}}^2
			\bigg)
\\&\quad
		- \CA\bigg(
			R(y_{13},y_{23})
			+ \frac{1}{2} \Log{y_{13}}^2
			+ \frac{1}{2} \Log{y_{23}}^2
			\bigg)
		- 2 \CF
		+ (2\CF + \CA)\frac{\pi^2}{2}
\\&\quad
		+ \bigg(
		3\CF + \frac{11\CA}{6} - \frac{2\Nf \TR}{3}
		\bigg) L \bigg]
\\&
	+ 8 \pi \frac{\as}{\Fep} \mu^{2\ep} 
	\SME{\bbar}{(0)}{ } 
	(\CA - \CF) \CF 
	\frac{1}{m_H^2}
	\bigg(\frac{1}{y_{13}} + \frac{1}{y_{23}}	\bigg)
	+ \Oe{}\bigg\}
\,,
\esp
\eeq
where
\beq
R(x,y) = \Li{2}{1-x} + \Li{2}{1-y} + \Log{x} \Log{y} - \frac{\pi^2}{6}\,.
\eeq
%

%
%

\subsection{Four partons}

In our computation we need the $H \to$ four partons squared matrix
elements at tree level in $d=4$ dimensions. We checked our formulae,
presented below, with {\sc GoSam} \cite{Cullen:2011ac,Cullen:2014yla}.

For $H \to \bbar q\qb$ we have
\beq
\SME{\bbar q\qb}{(0)}{} = 
	\Big(8 \pi \as \mu^{2\ep}\Big)^2
	\SME{\bbar}{(0)}{ } 
	\frac{1}{m_H^4} 
	\Big[C_{\bbar q\qb}(p_1,p_2,p_3,p_4) \TR \CF \Big]
	+ \Oe{}\,,
\eeq
where
\beq
\bsp
C_{\bbar q\qb}(p_1,p_2,p_3,p_4) =& 
	\bigg[\frac{1}{2 y_{34}}
	- \frac{1}{2 y_{134}}
	- \frac{1}{2 y_{134}^2}
	- \frac{1+y_{13}}{y_{134} y_{34}} 
	+ \frac{1+4 y_{13}+y_{34}}{2 y_{134} y_{234}} 
	- \frac{y_{13}}{y_{134}^2 y_{34}} 
\\&
	+ \frac{1+2 y_{13}+2 y_{13}^2+2 y_{13} y_{23}}{2 y_{134} y_{234} y_{34}} 
	- \frac{y_{13}^2}{y_{134}^2 y_{34}^2} 
	+ \frac{y_{13} y_{23}}{y_{134} y_{234} y_{34}^2}
\\&
	+ (1 \leftrightarrow 2)
	+ (3 \leftrightarrow 4)
	+ (1 \leftrightarrow 2\,,\, 3 \leftrightarrow 4)
	\bigg]
\,.
\esp
\eeq

For $H \to \bbar \bbar$ we find
\beq
\bsp
\SME{\bbar \bbar}{(0)}{} = 
	\Big(8 \pi \as \mu^{2\ep}\Big)^2
	\SME{\bbar}{(0)}{ } 
	\frac{1}{m_H^4} 
	&\Big[
	A_{\bbar \bbar}(p_1,p_2,p_3,p_4) \CA \CF
	+ B_{\bbar \bbar}(p_1,p_2,p_3,p_4) \CF^2
\\&
	+ C_{\bbar \bbar}(p_1,p_2,p_3,p_4) \TR \CF 
	\Big]
	+ \Oe{}\,,
\esp
\eeq
where
\beq
\bsp
A_{\bbar \bbar}(p_1,p_2,p_3,p_4) =&
	\bigg[\frac{1}{2 y_{12}} 
	- \frac{1}{2 y_{123}} 
	- \frac{1}{2 y_{124}} 
	+ \frac{y_{23}+y_{24}}{y_{12} y_{14}} 
	+ \frac{y_{13}+y_{14}}{y_{12} y_{23}} 
\\&
	- \frac{4 y_{13}-3 y_{14}+y_{24}-3 y_{34}}{4 y_{12} y_{123}} 
	- \frac{y_{13}-3 y_{23}+4 y_{24}-3 y_{34}}{4 y_{12} y_{124}} 
\\&
	+ \frac{y_{13}-4 y_{23}-3 y_{24}-2 y_{34}}{2 y_{12} y_{134}} 
	- \frac{3 y_{13}+4 y_{14}-y_{24}+2 y_{34}}{2 y_{12} y_{234}} 
\\&
	- \frac{2 y_{12}-3 y_{13}-y_{14}-y_{23}-3 y_{24}-8 y_{34}}{4 y_{123} 
		y_{124}} 
	+ \frac{3 y_{12}+y_{24}}{2 y_{123} y_{134}} 
\\&
	+ \frac{3 y_{12}+y_{13}}{2 y_{124} y_{234}} 
	- \frac{y_{13} (y_{14}+y_{24}+y_{34})}{y_{12} y_{123}^2} 
	- \frac{y_{24} (y_{13}+y_{23}+y_{34})}{y_{12} y_{124}^2} 
	+ \frac{y_{34} (y_{14}+y_{23})}{y_{12} y_{123} y_{124}} 
\\&
	+ \frac{2 y_{13}^2-2 y_{13} y_{24}-2 y_{13} y_{34}-2 y_{23}^2
		-4 y_{23} y_{24}-2 y_{23} y_{34}+y_{24}^2+2 y_{24} y_{34}
		+2 y_{34}^2}{4 y_{12} y_{123} y_{14}} 
\\&
	+ \frac{y_{13}^2-4 y_{13} y_{14}-2 y_{13} y_{24}+2 y_{13} y_{34}
		-2 y_{14}^2-2 y_{14} y_{34}+2 y_{24}^2-2 y_{24} y_{34}
		+2 y_{34}^2}{4 y_{12} y_{124} y_{23}} 
\\&
	- \frac{y_{13}^2+y_{13} y_{34}+y_{14}^2+2 y_{14} y_{24}
		+3 y_{14} y_{34}-2 y_{23}^2+y_{24}^2+4 y_{24} y_{34}
		+4 y_{34}^2}{4 y_{12} y_{123} y_{134}} 
\\&
	- \frac{2 y_{14}^2-2 y_{14} y_{23}+2 y_{14} y_{24}+2 y_{14} y_{34}
		+y_{23}^2+y_{24}^2+2 y_{24} y_{34}+3y_{34}^2}{4 y_{12} y_{123} 
		y_{234}}
\\&
  	- \frac{y_{13}^2+2 y_{13} y_{23}+2 y_{13} y_{34}+y_{14}^2
		-2 y_{14} y_{23}+2 y_{23}^2+2 y_{23} y_{34}+3 y_{34}^2}{4 y_{12} 
		y_{124} y_{134}} 
\\&
	- \frac{y_{13}^2+2 y_{13} y_{23}+4 y_{13} y_{34}-2 y_{14}^2
		+y_{23}^2+3 y_{23} y_{34}+y_{24}^2+y_{24} y_{34}
		+4 y_{34}^2}{4 y_{12} y_{124} y_{234}} 
\\&
	- \frac{2 y_{23}^3+2 y_{23}^2 y_{24}+y_{23} y_{24}^2}{4 y_{12} 
		y_{123} y_{134} y_{14}} 
	- \frac{y_{13}^2 y_{14}+2 y_{13} y_{14}^2+2 y_{14}^3}{4 y_{12} 
		y_{124} y_{23} y_{234}}
\\&
	+ (1 \leftrightarrow 3)
	+ (2 \leftrightarrow 4)
	+ (1 \leftrightarrow 3\,,\, 2 \leftrightarrow 4)
	\bigg]
\,,
\esp
\eeq
while
\beq
B_{\bbar \bbar}(p_1,p_2,p_3,p_4) = -2 A_{\bbar \bbar}(p_1,p_2,p_3,p_4)
\eeq
and finally 
\beq
C_{\bbar \bbar} = 
	\Big[C_{\bbar q\qb}(p_1,p_2,p_3,p_4)
	+ (1 \leftrightarrow 3)
	+ (2 \leftrightarrow 4)
	+ (1 \leftrightarrow 3\,,\, 2 \leftrightarrow 4)\Big]\,.
\eeq

For $H \to \bbar gg$ we obtained:
\beq
\bsp
\SME{\bbar gg}{(0)}{} = 
	\Big(8 \pi \as \mu^{2\ep}\Big)^2
	\SME{\bbar}{(0)}{ } 
	\frac{1}{m_H^4} 
	\Big[&
	A_{\bbar gg}(p_1,p_2,p_3,p_4) \CA \CF
	+ B_{\bbar gg}(p_1,p_2,p_3,p_4) \CF^2
	\Big]
	+ \Oe{}\,,
\esp
\eeq
where
\beq
\bsp
A_{\bbar gg}(p_1,p_2,p_3,p_4) =&
	\bigg[\frac{7}{2 y_{13}} 
	+ \frac{5}{4 y_{134}} 
	+ \frac{1}{2 y_{134}^2} 
	-\frac{3 (1-y_{23}-y_{34})}{2 y_{13} y_{14}} 
	-\frac{3 (2-2 y_{14}-y_{34})}{2 y_{13} y_{23}}
\\& 
	- \frac{8-10 y_{14}-7 y_{34}}{4 y_{13} y_{24}} 
	- \frac{3 (2-2 y_{14}-y_{23}-y_{24})}{4 y_{13} y_{34}} 
	+ \frac{3+y_{23}-y_{24}+2 y_{34}}{4 y_{13} y_{134}} 
\\&
	+ \frac{10-4 y_{14}+3 y_{23}-y_{24}+4 y_{34}}{4 y_{13} y_{234}} 
	+ \frac{2+y_{13}}{y_{134} y_{34}} 
	- \frac{8+8 y_{13}+5 y_{34}}{4 y_{134} y_{234}} 
\\&
	+ \frac{y_{13}}{y_{134}^2 y_{34}} 
	+ \frac{4-3 y_{24}-6 y_{34}+y_{24}^2+3 y_{24} y_{34}
		+3 y_{34}^2}{2 y_{13} y_{14} y_{23}} 
\\&
	+ \frac{2-4 y_{14}+2 y_{14}^2+2 y_{14} y_{23}}{4 y_{13} y_{24} y_{34}} 
	+ \frac{4-3 y_{24}+3 y_{34}+y_{24}^2-y_{24} y_{34}
		+y_{34}^2}{2 y_{13} y_{134} y_{23}}
\\& 
	+ \frac{4-4 y_{14}+2 y_{23}-2 y_{24}+2 y_{14}^2-2 y_{14} y_{23}
		+2 y_{14} y_{24}+y_{23}^2+y_{24}^2}{4 y_{13} y_{234} y_{34}}
\\& 
	- \frac{8+3 y_{23}-3y_{24}+9 y_{34}+y_{23}^2+3 y_{23} y_{34}
		+y_{24}^2-y_{24} y_{34}+4 y_{34}^2}{4 y_{13} y_{134} y_{234}} 
\\&
	- \frac{2+y_{13}+y_{13}^2+y_{13} y_{23}}{y_{134} y_{234} y_{34}} 
	+ \frac{y_{13}^2}{y_{134}^2 y_{34}^2} 
	- \frac{2-4 y_{34}+3 y_{34}^2-y_{34}^3}{8 y_{13} y_{14} y_{23} y_{24}} 
\\&
	- \frac{y_{13} y_{23}}{y_{134} y_{234} y_{34}^2} 
	- \frac{2+4 y_{34}+3 y_{34}^2+y_{34}^3}{4 y_{13} y_{134} y_{23} y_{234}}
\\&
	+ (1 \leftrightarrow 2)
	+ (3 \leftrightarrow 4)
	+ (1 \leftrightarrow 2\,,\, 3 \leftrightarrow 4)
	\bigg]
\esp
\eeq
and
\beq
\bsp
B_{\bbar gg}(p_1,p_2,p_3,p_4) =&
	\bigg[-\frac{11}{2 y_{13}} 
	+ \frac{1}{2 y_{134}^2} 
	+ \frac{3 (1-y_{23}-y_{34})}{y_{13} y_{14}} 
	+ \frac{3 (2-2 y_{14}-y_{34})}{y_{13} y_{23}} 
\\&	
	+ \frac{7-6 y_{14}-6 y_{34}}{2 y_{13} y_{24}} 
	+ \frac{1-y_{34}}{2 y_{13} y_{134}} 
	- \frac{5-4 y_{14}+y_{23}-y_{24}+3 y_{34}}{2 y_{13} y_{234}} 
	+ \frac{1+y_{34}}{y_{134} y_{234}} 
\\& 
	- \frac{y_{14}-y_{34}}{2 y_{13} y_{134}^2}
	- \frac{4-3 y_{14}-6 y_{34}+y_{14}^2+3 y_{14} y_{34}
		+3 y_{34}^2}{y_{13} y_{23} y_{24}} 
\\&
	- \frac{4-3 y_{24}+3 y_{34}+y_{24}^2-y_{24} y_{34}
		+y_{34}^2}{y_{13} y_{134} y_{23}} 
	- \frac{4+y_{14}-2 y_{23}-y_{34}}{2 y_{13} y_{134} y_{24}} 
\\&
	+ \frac{y_{34} (6+y_{23}-y_{24}+3 y_{34})}{2 y_{13} y_{134} y_{234}} 
	+ \frac{2-4 y_{34}+3 y_{34}^2-y_{34}^3}{4 y_{13} y_{14} y_{23} y_{24}}
	+ \frac{2+4 y_{34}+3 y_{34}^2+y_{34}^3}{2 y_{13} y_{134} y_{23} y_{234}} 
\\& 
	+ \frac{1}{y_{13} y_{134} y_{234} y_{24}}
	+ (1 \leftrightarrow 2)
	+ (3 \leftrightarrow 4)
	+ (1 \leftrightarrow 2\,,\, 3 \leftrightarrow 4)
	\bigg]
\,.
\esp
\eeq


\section{$\bI_1^{(0)}$ insertion operator to $\Oe{}$}
\label{appx:I103j}

We present the $\bI_1^{(0)}(\{p\}_m;\ep)$ insertion operator in \eqn{eq:I10} 
to $\Oe{}$. More precisely, we give the $\ep$-expansion of the kinematic 
functions $\IcC{1,i}{(0)}(x,\ep)$ and $\IcS{1}{(0),(i,k)}(Y,\ep)$ which appear 
in \eqn{eq:I10} up to and including finite terms.

Starting with $\IcC{1,i}{(0)}(x,\ep)$, we have
\bal
\IcC{1,q}{(0)}(x,\ep) &= 
	[\IcC{ir}{(0)}]_{qg}(x,\ep) 
	- [\IcC{ir}{}\IcS{r}{(0)}](\ep) \,,
\\
\IcC{1,g}{(0)}(x,\ep) &= 
	\frac{1}{2}[\IcC{ir}{(0)}]_{gg}(x,\ep)  + \Nf [\IcC{ir}{(0)}]_{q\qb}(x,\ep)  
	- [\IcC{ir}{}\IcS{r}{(0)}](\ep)\,,
\eal
where
\bal
&
[\IcC{ir}{(0)}]_{qg}(x, \ep) = 
	\frac{1}{\ep^2}
	+\left(\frac{3}{2}-2 \ln(x)\right)\frac{1}{\ep}
\nt\\&\quad
	+ 2\left(1+\frac{1}{(1-x)^5}\right) \Li{2}{1-x} -\frac{\pi ^2}{2} +2 \ln^2(x)
\nt\\&\quad
	+\left(\frac{8}{3 (1-x)^5} - \frac{3}{2 (1-x)^4} - \frac{1}{3 (1-x)^3} 
	+ \frac{1}{3 (1-x)^2} + \frac{3}{2 (1-x)} -\frac{17}{3}\right) \ln(x)
\nt\\&\quad
	+\frac{2}{3 (1-x)^4}-\frac{2}{3 (1-x)^3}-\frac{5}{12 (1-x)^2}+\frac{5}{24 (1-x)} 
	+\frac{89}{24}
	+\Oe{}\,,
\\[1em]
&
[\IcC{ir}{(0)}]_{q\qb}(x, \ep) = 
	\frac{\TR}{\CA} \bigg\{-\frac2{3 \ep}
\nt\\& \quad
	+ \frac{2 }{3 } \left(1+\frac{1}{(1-x)^5}\right) \ln(x) - \frac{160}{3} 
	\left(\frac{2}{(2-x)^6}-\frac1{(2-x)^5}\right) \ln\left(\frac{x}{2}\right)
\nt\\& \quad
	+\frac{2 }{3  (1-x)^4}+\frac{1}{3  (1-x)^3}+\frac{2 }{9  (1-x)^2}
	+\frac{1}{6  (1-x)}-\frac{5 }{2 }
\nt\\& \quad
	-\frac{160 }{3  (2-x)^5}+\frac{40 }{3  (2-x)^4}+\frac{20 }{9  (2-x)^3}
	+\frac{5 }{9  (2-x)^2}+\frac{1}{6  (2-x)}
	\bigg\}
	+\Oe{}\,,
\\[1em]
&
[\IcC{ir}{(0)}]_{gg}(x, \ep) = 
	\frac{2}{\ep^2}+ \left(\frac{11}{3}-4 \ln(x)\right)\frac{1}{\ep}
\nt\\& \quad
	+4 \left(1+\frac{1}{(1-x)^5}\right) \Li{2}{1-x} +4 \ln^2(x) -\pi ^2
	+ \frac{160}{3}  \left(\frac{2}{(2-x)^6}-\frac1{(2-x)^5}\right) 
	\ln\left(\frac{x}{2}\right)
\nt\\& \quad
	+\left(\frac{14}{3 (1-x)^5}-\frac{3}{(1-x)^4}-\frac{2}{3 (1-x)^3}
	+\frac{2}{3 (1-x)^2}+\frac{3}{1-x}-12\right) \ln(x)
\nt\\& \quad
	+\frac{2}{3 (1-x)^4}-\frac{5}{3 (1-x)^3}-\frac{19}{18 (1-x)^2}
	+\frac{1}{4 (1-x)} +\frac{37}{4}
\nt\\& \quad
	+\frac{160}{3 (2-x)^5} -\frac{40}{3 (2-x)^4}-\frac{20}{9 (2-x)^3}
	-\frac{5}{9 (2-x)^2}-\frac{1}{6 (2-x)}
	+\Oe{}\,,
\eal
and

\beq
[\IcC{ir}{}\IcS{r}{(0)}](\ep) =  
	\frac{1}{\ep^2} + \frac{11}{3\ep}
	- \frac{7}{6} \pi ^2 + \frac{329}{18}
	+\Oe{}\,.
\eeq

Turning to $\IcS{1}{(0),(i,k)}(Y,\ep)$, we have simply
\beq
\IcS{1}{(0),(i,k)}(Y,\ep) = 
	[\IcS{r}{(0)}]^{(i,k)}(Y,\ep)\,,
\eeq
where
\beq
\bsp
&
[\IcS{r}{(0)}]^{(i,k)}(Y, \ep) = 
	- \frac{1}{\ep^2} 
   	+ \left( \ln(Y) -\frac{11}{3} \right) \frac{1}{\ep}
   	-\Li{2}{1-Y} 
\\& \quad			
	-\frac1{2} \ln^2(Y) 
	+ \frac{7}{6}\pi ^2 
	+ \frac{11}{3} \ln(Y) 
   	-\frac{317}{18}
	+\Oe{}\,.
\esp
\eeq

\newpage 



\providecommand{\href}[2]{#2}\begingroup\raggedright\endgroup


\end{document}